\documentclass[twocolumn]{aastex6}
\usepackage[]{units}
\usepackage[update,prepend]{epstopdf}
\usepackage{graphicx}
\usepackage{amssymb}
\usepackage{amsmath}
\usepackage{threeparttable}
\usepackage{hyperref}
\usepackage{color}

\shorttitle{Spectropolarimetry of Obscured Quasars}
\shortauthors{Alexandroff et al.}

\begin{document}
\title{Spectropolarimetry of High Redshift Obscured and Red Quasars} 

\author{Rachael M. Alexandroff\altaffilmark{1}, Nadia L. Zakamska\altaffilmark{1,2}, Aaron J. Barth\altaffilmark{3}, Fred Hamann\altaffilmark{4}, Michael A. Strauss\altaffilmark{5}, Julian Krolik\altaffilmark{1}, Jenny E. Greene\altaffilmark{5}, Isabelle P\^{a}ris\altaffilmark{6}, Nicholas P. Ross\altaffilmark{7}}
\altaffiltext{1}{Center for Astrophysical Sciences, Department of Physics and Astronomy, Johns Hopkins University, Baltimore, MD 21218, USA}
\altaffiltext{2}{Deborah Lunder and Alan Ezekowitz Founders' Circle Member, Institute for Advanced Study, Einstein Dr., Princeton, NJ 08540, USA}
\altaffiltext{3}{Department of Physics and Astronomy, University of California, Irvine, 4129 Frederick Reines Hall, Irvine, CA 92697, USA}
\altaffiltext{4}{Department of Physics \& Astronomy, University of California, Riverside, CA 92507, USA}
\altaffiltext{5}{Department of Astrophysical Sciences, Princeton University, Princeton, NJ 08544, USA}
\altaffiltext{6}{Aix Marseille Universit\'{e}, CNRS, LAM, Laboratoire d'Astrophysique de Marseille, 13013 Marseille, France}
\altaffiltext{7}{Institute for Astronomy, University of Edinburgh, Royal Observatory, Blackford Hill, Edinburgh EH9 3HJ, United Kingdom}
\email{rachael.alexandroff@dunlap.utoronto.ca}

\label{firstpage}

\begin{abstract}
Spectropolarimetry is a powerful technique that has provided critical support for the geometric unification model of local active galactic nuclei.  In this paper, we present optical (rest-frame UV) Keck spectropolarimetry of five luminous obscured (Type 2) and extremely red quasars (ERQs) at $z \simeq 2.5$. Three objects reach polarization fractions of $\gtrsim 10\%$ in the continuum. We propose a model in which dust scattering is the dominant scattering and polarization mechanism in our targets, though electron scattering cannot be completely excluded. Emission lines are polarized at a lower level than is the continuum. This suggests that the emission-line region exists on similar spatial scales as the scattering region. In three objects we detect an intriguing 90 degree swing in the polarization position angle as a function of line-of-sight velocity in the emission lines of Ly$\alpha$, \ion{C}{4} and \ion{N}{5}. We interpret this phenomenon in the framework of a geometric model with an equatorial dusty scattering region in which the material is outflowing at several thousand km sec$^{-1}$. Emission lines may also be scattered by dust or resonantly. This model explains several salient features of observations by scattering on scales of a few tens of pc. Our observations provide a tantalizing view of the inner region geometry and kinematics of high-redshift obscured and extremely red quasars. Our data and modeling lend strong support for toroidal obscuration and powerful outflows on the scales of the UV emission-line region, in addition to the larger scale outflows inferred previously from the optical emission-line kinematics. 
\end{abstract}


\section{Introduction}
\label{sec:intro}

The classical unification model of Active Galactic Nuclei
\citep[AGN;][]{Antonucci1993,Urry1995} explains the difference between
unobscured or ``Type 1" AGN and obscured or ``Type 2" AGN as the
result of differences in viewing angle.  In this model, a
geometrically and optically thick torus of gas and dust surrounds the
AGN accretion disk and broad-line region (BLR).  Therefore, if the
viewer's line of sight points through the torus, most of the quasar emission from the accretion disk and the BLR in the optical, ultraviolet (UV) and soft X-rays is obscured. Such an obscured AGN is classified as a ``Type 2" object and is characterized at optical and ultraviolet wavelengths by little to no continuum emission and an absence of broad lines \citep[e.g.][]{khac74, Kauffmann2003,Hao2005}. Quasars are high-luminosity AGN (typical bolometric luminosities $\gtrsim 10^{45}$ erg s$^{-1}$) and it is only in the last fifteen years or so that appreciable numbers of Type 2 quasars have been identified, and the unification model has been extended to high AGN luminosities \citep{Norman2002,Stern2002,Zakamska2003,Stern2005,Reyes2008,Treister2009,Donley2012,Alexandroff2013,Yuan2016}.  

In the local universe, a convincing case for the geometric unification model was first made using optical spectropolarimetry of nearby Type 2 AGN \citep{Antonucci1985,Miller1991}. Though a direct line of sight to the optical AGN emission is blocked in the geometry of Type 2 AGN, this emission can still escape along the unobscured direction perpendicular to the torus and scatter off of free electrons or dust particles in the quasar host galaxy and into the observer's line of sight. This scattering causes the emission to become linearly polarized, making optical polarimetry and spectropolarimetry the best way to view this scattered emission from the nucleus \citep{Miller1990,Tran1995a}. 

This process is presumably also occurring in unobscured or Type 1 AGN, but linearly polarized optical emission represents a much smaller fraction of the total light in these objects and the geometry is less favourable \citep[e.g.][]{Borguet2008}. In fact, typical levels of polarization in unobscured AGN are only 0.5\% \citep{Berriman1990}.  In contrast, optical polarization levels in Type 2 quasars at low redshift can reach values of at least a few percent \citep{Smith2002,Smith2003,Zakamska2005} and occasionally as high as $\gtrsim$ 20\% \citep{Hines1995,Smith2000,Zakamska2005}. In addition, light from the BLR scatters into our line of sight by the same process, so the spectrum of a Type 2 AGN in polarized light possesses broad emission lines \citep[e.g.][]{Antonucci1985, Zakamska2005}. Thus, high measured levels of polarization and broad emission lines seen only in polarization are tell-tale signs of obscured active nuclei in the classical orientation model.  Indeed, it was imaging polarimetry and later deep spectropolarimetry that confirmed the presence of hidden AGN at the centers of radio galaxies-- demonstrating that these objects were in fact radio loud Type 2 AGN \citep[see][and references therein]{Antonucci1993,Vernet2001,Tadhunter2005}. 

Few, if any, spectropolarimetric observations of radio-quiet obscured quasars at the peak of galaxy formation ($z\sim2.5$) are available because until recently, such objects could not be readily identified. Furthermore, obscured quasars are by definition very faint at rest-frame optical and UV wavelengths, whereas spectropolarimetry requires high signal-to-noise ratio observations to detect polarized flux at the level of a few percent. Therefore, we can only perform polarimetry and especially spectropolarimetry of high-redshift obscured quasars with large telescopes and / or long integration times.  \citet{Alexandroff2013} reported the results of spectropolarimetry performed on two high redshift ($z \sim 2.5$) rest-frame UV-selected Type 2 quasar candidates using the CCD spectropolarimeter \citep[SPOL;][]{Schmidt1992b} at the 6.5m MMT telescope. They demonstrated the presence of a polarization signature at the level of a few percent, but low signal-to-noise (S/N) ratios and availability of data for only two objects made further interpretation difficult. 

While it is clear what polarization signal is expected from a quasar
obscured by a classical torus (Type 2), perhaps not all quasars
possess a classical obscuring region. For example, \citet{Sanders1988}
and \citet{Hopkins2006} argue that some obscured quasars may represent
a particular phase in quasar evolution after a merger or an
instability within the galactic disk enshrouds the object in gas and
dust. If this is the case, obscuration may originate on galaxy-wide
scales. We then might expect a population of objects in which the
openings in the obscured material does not give rise to well-organized conical emission.  Red quasars at both low \citep[e.g.][]{Smith2000,Glikman2007,Urrutia2009,Glikman2012,Glikman2013} and high \citep{Eisenhardt2012,Wu2012,Tsai2015,Banerji2015,Ross2015,Hamann2017} redshift may be undergoing a phase of quasar evolution where the obscuring region is not a classical torus as evidenced by the presence of galaxy-wide outflows \citep{Zakamska2016} and merger signatures \citep{Glikman2015}.  Such objects would not have symmetric scattering regions (an example source was presented by \citealt{Schmidt2007}), and thus would have a lower net polarization fraction than classical Type 2 quasars. 

In this work we use spectropolarimetry to probe the geometry of
obscuration of high-redshift Type 2 quasars and extremely red quasars
(ERQs). We present observations of five such objects conducted with
the Low Resolution Imaging Spectrometer \citep[LRIS;][]{Oke1995} in
polarimetry mode \citep{Goodrich1995a}. We describe our sample
selection as well as observations and data reduction in Section
\S~\ref{sec:sample}. Then Section \S~\ref{sec:analysis} describes our
data analysis and results. We introduce our proposed model and
discuss our results in Section \S~\ref{sec:discussion}. We conclude in
Section \S~\ref{sec:conclusions}.   

Throughout this paper, we adopt a cosmology with $h$ = 0.7, $\Omega_m$ = 0.3 and $\Omega_{\Lambda}$ = 0.7. We use SDSS Jhhmm+ddmm notation throughout the text (full coordinates are listed in Table \ref{tab:info}) with an additional marker letter to represent how the source was originally identified (see section \ref{ssec:selection} for more details). Throughout, $U$ and $Q$ refer to Stokes flux densities while $u$ and $q$ refer to Stokes parameters relative to the total continuum.

\section{Sample Selection, Observations and Data Reduction}
\label{sec:sample}

\subsection{Parent Sample}
\label{ssec:selection}

Our sample consists of five obscured quasars selected using two distinct methods: two Type 2 quasar candidates at redshift $z \sim 2.5$ \citep{Alexandroff2013}, two extremely red quasars (ERQs) at the same redshift \citep{Ross2015, Hamann2017}, and one object which meets the criteria of both selection methods (Table \ref{tab:info}). We briefly describe the parent sample selection methods below. We were particularly interested in searching for evidence of differences in the covering factor of obscuration (as traced by the polarization fraction) between these two populations.

Two of our sources, which we mark with a `T' at the end of the designation, SDSSJ1515+1757T and SDSSJ1623+3122TE, were originally selected from the Sloan Digital Sky Survey (SDSS) Baryon Oscillation Spectroscopic Survey \citep[BOSS;][]{Dawson2013} by their narrow line widths (FWHM $< 2000$ km s$^{-1}$ in both \ion{C}{4} and Lyman $\alpha$) and weak continuum in the rest-frame UV, to form an optically-selected sample of high redshift Type 2 quasar candidates \citep[for more details see][]{Alexandroff2013}.  Only objects from Data Release 9 \citep{Ahn2012} or earlier were included in this search, which yielded a sample of 145 Type 2 quasar candidates. 

An additional three sources, labeled with an `E', SDSSJ1232+0912E,
SDSS1652+1728E and SDSSJ2215$-$0056E were selected based upon a
combination of data from the SDSS and the Wide-Field Infrared Survey
Explorer \citep[WISE;][]{White2010} AllWISE data
release\footnote{http://irsa.ipac.caltech.edu/}
\citep{Ross2015,Hamann2017}. This selection targeted objects with high
infrared-to-optical ratios expected in obscured quasars. A sub-sample
of these red objects show unusual spectroscopic properties, including
large rest equivalent width (REW) emission lines ($\gtrsim 100$ \AA),
unusually high \ion{N}{5}/Ly$\alpha$ ratios, and emission lines with
high kurtosis (lacking the typical broad wings of gaussian emission
lines). These objects were labelled Extremely Red Quasars (ERQs). A
final sample of 97 ERQs in the redshift range of $2.0 < z < 3.4$ was
selected to have a colour between the SDSS $i$-band and the WISE $W3$
band (12 $\mu$m) $\geq$ 4.6 mag, and a measured CIV REW $> 100$ \AA\ \citep{Hamann2017}. 

There is also some overlap between the two samples. While SDSSJ1652+1728E does not meet the strict cutoff required of the \citet{Alexandroff2013} sample of Type 2 candidates, we note its relatively narrow FWHM in \ion{C}{4} ($< 2500$ km s$^{-1}$). Additionally, SDSSJ1623+3122TE was primarily selected as a Type 2 quasar candidate, but is also listed as part of the ``expanded" sample of ERQs, defined to have ``ERQ-like" line properties even if they are less red than the principal sample in $i-W3$ colours \citep{Hamann2017}. We label it as TE in Table \ref{tab:info} to denote this joint classification. Both the \citet{Alexandroff2013} and the \citet{Hamann2017} methods aim to select obscured quasars at high redshifts. While we still do not fully understand the physical properties of each of these populations, SDSSJ1623+3122TE and SDSSJ1652+1728E demonstrate that there can be some overlap between the methods.

\begin{table*}[htp]
\caption{Type 2 quasars and extremely red quasars with spectropolarimetric observations \label{tab:info}}
\begin{center}
\begin{tabular}{lllcccccccc}
\hline
Short Name & RA & Dec & $z$ & $r$ &CIV& CIV&[OIII] & log & $i-W3$ & type\\
&&&$L_{\rm [OIII]}$,&mag&FWHM&$kt_{80}$&FWHM&&&\\
&hh:mm:ss&dd:mm:ss&&AB&km s$^{-1}$&&km s$^{-1}$&erg s$^{-1}$&AB&\\
\hline
SDSSJ1232+0912E & 12:32:41.73 & +09:12:09.37 & 2.374 &21.11$\pm$0.05&4787 $\pm$ 52&0.37&5627&43.92&6.8&ERQ \\
SDSSJ1515+1757T & 15:15:44.00 & +17:57:53.06 & 2.402 &22.24$\pm0.12$&1118 &--&856.4&43.98&2.0&T2\\
SDSSJ1623+3122TE & 16:23:27.66 & +31:22:04.29 & 2.344 &21.60$\pm$0.09&1572$\pm$68&0.34&658.3&44.08&4.5&T2E \\
SDSSJ1652+1728E & 16:52:02.64 & +17:28:52.38 & 2.942 &20.38$\pm$0.02&2403 $\pm$ 45&0.33&1461&44.87&5.4&ERQ \\
SDSSJ2215$-$0056E &22:15:24.00 & $-$00:56:43.80 & 2.493 &22.27$\pm$0.12&4280 $\pm$ 112&0.37&3057&43.64&6.2&ERQ \\
\hline
\end{tabular}
\end{center}
\begin{tablenotes}
\item[a] Type 2 quasars and ERQs presented in this paper. Our targets span a range of properties in their line widths and colours. Position and redshift information are from the SDSS Skyserver Data Release 12.  \ion{C}{4} FWHM, REW and $kt_{80}$ were measured by \citet{Hamann2017} for four objects. Measurements for SDSS1515+1757T are taken from the SDSS Skyserver Data Release 12. For SDSSJ1232+0912E and SDSSJ2215$-$0056E, [\ion{O}{3}] FWHM and luminosity are from \citet{Zakamska2016}, and for the remaining objects these measurements are from the Gemini North data presented in Section \ref{ssec:optical}.
\end{tablenotes}
\end{table*}%

\subsection{Rest-frame Optical Properties}
\label{ssec:optical}

SDSS covers the rest-frame UV at the redshift of our objects, and therefore follow-up near-infrared (NIR) spectroscopy is necessary to probe their rest-frame optical properties. For the purposes of the current study, we use the near-infrared (rest-frame optical) spectra primarily to distinguish between classical Type 1 and Type 2 objects and to look for outflow signatures. 

\begin{figure*}
\includegraphics[scale=0.8,trim=20 340 40 100,clip=true]{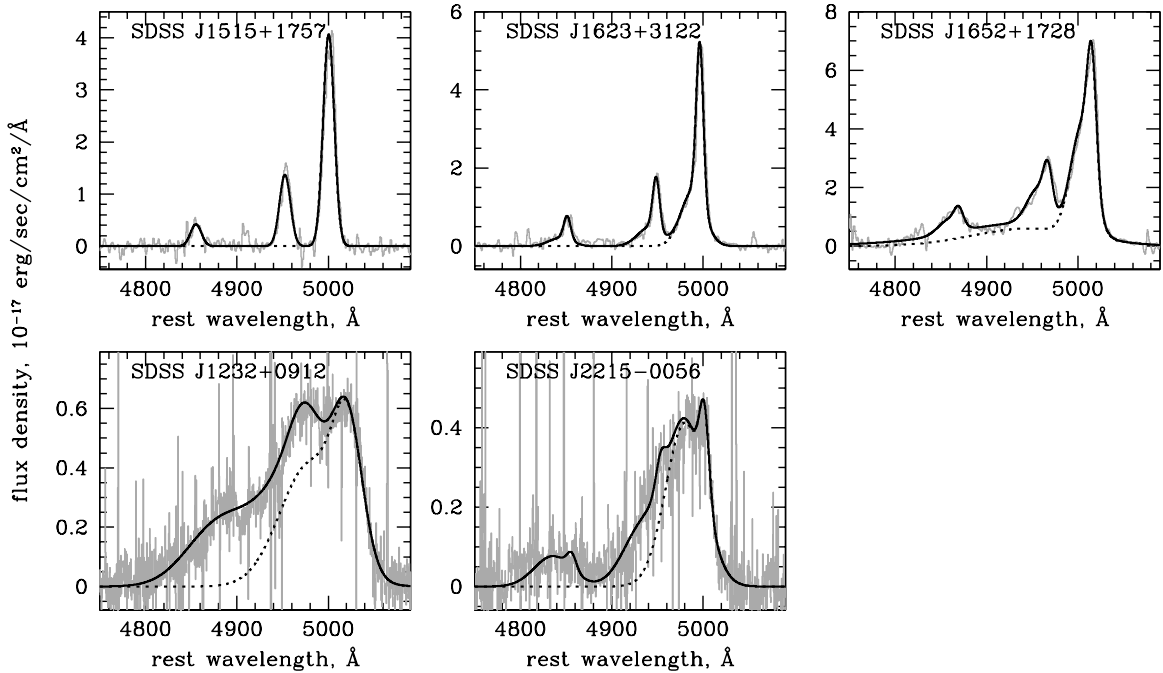}
\caption{\small The continuum-subtracted [\ion{O}{3}]+H$\beta$ complex and best fits for the five objects studied in this paper.  The spectra were obtained for the top three objects with Gemini GNIRS and have not previously been published.  Spectra for the bottom two objects were obtained with the VLT XShooter and were published in \citet{Zakamska2016}.  The grey histograms show the data, the solid black line shows the best overall fit and the dotted black line shows the [\ion{O}{3}]$\lambda$5007 \AA\ fitted profile separately. In SDSSJ1515+1757T, SDSSJ1623+3122TE and SDSSJ1652+1728E the best fits are kinematically tied (i.e.,  [\ion{O}{3}] and H$\beta$ have the same velocity structure), and the best velocity fit is comprised of one, two and three Gaussian components, respectively. The best fits for SDSSJ1232+0913E and SDSSJ2215$-$0056E were previously presented in \citet{Zakamska2016} and are reproduced here. In SDSSJ1232+0913E, the best fit is kinematically untied; in SDSSJ2215$-$0056E it is kinematically tied; in both cases, two Gaussian components are used for  [\ion{O}{3}].}
\label{fig:nir_spectra}
\end{figure*}

The [\ion{O}{3}]$\lambda$5007 \AA+H$\beta$ spectra for all five targets studied in this Keck program are shown in Figure \ref{fig:nir_spectra}, with three objects on top from our new program on the Gemini Near-Infrared Spectrograph (GNIRS) on Gemini-North and two objects on the bottom from the previously published XShooter program \citep{Zakamska2016}. Our new GNIRS data were obtained in semester 15A in queue mode (PI Alexandroff). The instrument was operated in cross-dispersed mode using the 1/32 mm grating and with a slit width of 0\farcs45 to provide full wavelength coverage from  0.9 - 2.5 $\mu$m, which covers H$\alpha$, H$\beta$ and [\ion{O}{3}] in the rest-frame of our objects. Each object was observed in queue mode for a total of 60 minutes in a series of nodded 300s exposures along the slit. Data reduction was performed using the XDGNIRS package v2.0 \citep{Mason2015}, a wrapper script for the PyRAF tasks in the Gemini IRAF package. The script was run in interactive mode and the data were examined at each step in the process. As our targets are faint point sources, we manually set the position of the aperture for the trace of each order rather than having the program find it automatically. 

Using the calibrated 1D spectra, we performed multi-Gaussian emission-line fits as described by \citet{Zakamska2016}. Briefly, [\ion{O}{3}]$\lambda$4959 \AA\ and [\ion{O}{3}]$\lambda$5007 \AA\ are assumed to have the same velocity structure and a fixed flux density amplitude ratio of 0.337. Beyond this assumption, our fits are either ``kinematically tied'' -- i.e., the same velocity structure is assumed for [\ion{O}{3}], and H$\beta$ -- or ``kinematically untied'', in which H$\beta$ has velocity structure which is different from [\ion{O}{3}]. Our null hypothesis is that in Type 2 quasars, [\ion{O}{3}] and H$\beta$ are kinematically tied and their typical ratio is 10:1, with the caveat that both of these characteristics can be affected by strong outflows \citep{Zakamska2014,Zakamska2016}.

Unlike low-redshift Type 2 quasars selected to have narrow Balmer lines, follow-up observations in the rest-frame optical \citep{Greene2014} of twenty-five objects from our parent sample of Type 2 quasar candidates showed that $\sim 90\%$ of objects displayed a broad H$\alpha$ or H$\beta$ line and all had intermediate values of extinction (0 mag $< A_{\rm v}<$ 2.2 mag) akin to Type 1.8/1.9 quasars. Yet, both SDSSJ1515+1757T and SDSSJ1623+3122TE -- selected based on their narrow high-equivalent-width ultraviolet lines \citep{Alexandroff2013} -- show classical Type 2 quasar features in the rest-frame optical: they have weak continua, narrow H$\beta$ that has the same velocity structure as [\ion{O}{3}], and [\ion{O}{3}]/H$\beta$ ratios of 9.9 and 6.9, respectively. These specific objects were chosen for spectropolarimetry follow-up because they are unambiguously high-redshift Type 2 quasars.  SDSSJ1515+1757T shows no sign of [\ion{O}{3}] outflows, whereas SDSSJ1623+3122T has a noticeable blueshifted component in its [\ion{O}{3}] emission.  

SDSSJ1652+1728E has a mix of features. While it is best fitted with a kinematically tied model with [\ion{O}{3}]/H$\beta$=7.7, the rest-frame optical continuum is relatively strong and marginally inconsistent with a Type 2 classification.  In addition, the rest equivalent width of [\ion{O}{3}]$\lambda$5007 \AA~of 213 \AA~is intermediate between Type 1s and Type 2s at these luminosities \citep{Greene2014}. This object shows a strongly kinematically disturbed [\ion{O}{3}], with the width containing 80\% of the line power of $w_{80}=1760$ km sec$^{-1}$, which is in the upper $\sim 5\%$ of [\ion{O}{3}] velocity width in the $z<1$ Type 2 population \citep{Zakamska2014}. 

The remaining two targets, SDSSJ1232+0912E and SDSSJ2215$-$0056E, are ERQs. In the rest-frame optical, they have a mix of Type 2 and Type 1 characteristics as described in detail in \citet{Zakamska2016}.  The rest-frame optical spectra for SDSSJ1232+0912E is fitted with a kinematically untied model while in the case of SDSSJ2215$-$0056E there is no preference between the kinematically tied and untied ones.  The REW of [\ion{O}{3}]$\lambda$5007 \AA~for both is intermediate between expected Type 1 and Type 2 values.  Both objects have extremely fast [\ion{O}{3}] outflows-- the FWHMs are 5630 and 3060 km s$^{-1}$ respectively. Because [\ion{O}{3}] has a relatively low critical density, it must originate on relatively large scales, $\gg 100$ pc \citep{hama11, Zakamska2016}. The [\ion{O}{3}] velocities seen in ERQs are too large to be contained by any reasonable galaxy potential and perhaps indicate that these quasars are in the predicted~``blowout'' phase of quasar activity \citep{Hopkins2006}. 

The five objects presented here were selected for follow-up with Keck LRISp to be either Type 2 quasars with classifications confirmed with near-infrared spectroscopy, or to be ERQs with strong signs of galaxy-scale outflows. Thus, this sample allows us to probe the obscuring geometry and scattering efficiency of obscured quasars selected by two different methods.

\subsection{Observations and Data Reduction}

Data for this project were obtained as part of a NASA time allocation on the Keck Telescope in semester 2015A.  All observations were conducted over the course of a single night (UT 2015-05-22) on Keck I from Keck Headquarters in Waimea in clear conditions with seeing around 0\farcs6.  The observational setup was chosen to maximize S/N over resolution for our faint targets and to put the relevant emission lines of Ly$\alpha$ and \ion{C}{4} on the blue side CCD where cosmic ray (CR) effects are less strong than on the red side. We used a 1\farcs0 slit width and the D68 dichroic to separate the light into blue and red beams. The blue side spectrograph used a 300 grooves mm$^{-1}$ grism giving a dispersion of 1.43 \AA~pixel$^{-1}$, while the red side used a 400 grooves mm$^{-1}$ grating which gave a dispersion of 1.16 \AA~ pixel$^{-1}$. This blue side grism provides some level of contamination on the red end due to second order light at $\lambda \gtrsim 6400$ \AA. No binning either spatially or spectrally was applied during observations. The combined wavelength coverage from both the blue and red sides stretched from 3100 \AA\ to 10,300 \AA, with a small overlap at $\sim 6500-6800$ \AA. Calibration darks, dome flats and arcs were taken the afternoon before the start of observations, with additional dome flats taken at the end of the evening.

All science targets were observed in a standard sequence of exposures with four positions of the rotatable half-wave plate (0$^{\circ}$, 45$^{\circ}$, 22.5$^{\circ}$, and 67.5$^{\circ}$) lasting 600 seconds on the blue side and 520 seconds on the red side to synchronize the ends of the exposures due to the differing read-out times. Four quasars were observed for two full observation sequences for a total observed time of 4800.0/4160.0 seconds on the blue and red sides respectively. SDSSJ2215$-$0056E was observed for a single full sequence -- half the exposure of the other objects -- due to timing constraints. In addition, we observed the flux standard star Feige 34, the null (unpolarized) standard stars HD 109055 and BD+28$^{\circ}$4211 (also a flux standard) and the polarized standards HD 155528 and HD 204827.  All observations were done at the parallactic angle except for the special case of SDSS1652+1728E described further below.

CR hits were removed from the raw blue-side science images using the
IRAF imaging version of L.A.COSMIC \citep{VanDokkum2001}. The red
detectors are 300$\mu$m thick LBNL CCDs \citep{Rockosi2010}, in which
a single CR hit gives rise to an extended trail spanning many more
individual pixels than on the blue-side CCDs. Because each Stokes parameter is measured
using four separate beam spectra and differencing pairs of spectra,
any error in cleaning cosmic rays is multiplied four-fold for a single
Stokes parameter, and eight-fold for the polarization fraction or
polarization position angle.
Thus, the red side CCDs suffer so acutely from CR
contamination that the data were unusable. 

Data reduction and calibration were conducted following the methods of \citet{Miller1988} and \citet{Barth1999}. In brief, bias subtraction, flat fielding, spectral extraction, wavelength and flux calibration were first performed using IRAF, and polarization measurements were performed using routines written in IDL. Uncertainties due to photon-counting statistics and detector readout noise were propagated at every step of the reduction process, yielding error spectra for the Stokes parameters $Q$ and $U$. Extraction regions were 24 pixels wide (corresponding to 3\farcs24) except for SDSS1652+1728E which had a different spatial profile described below. The background region was calculated over 10 pixels (corresponding to 1\farcs35) above and below the extraction region and an additional 10 pixels away from the extraction region.  A wide extraction region was used because the science targets were very faint. 

SDSS1652+1728E is point-like in the SDSS images.  There is another point-like source 1\farcs6 away which does not have SDSS spectroscopy. During the observations, we placed the slit to cover both objects and to determine whether the second object might be physically related. The spectra of this companion showed no features and upon further inspection, its SDSS photometric colours ($u-g = 0.97$, $g-i = 0.58$) place it in the stellar locus for SDSS point-like objects \citep{Yeche2010}. The presence of this likely foreground star in the slit makes proper extraction of SDSS1652+1728E complicated as the sources are so close as to be blended in the 2D spectra. We moved the background region an additional ten pixels away and made the extraction region only 20 pixels to avoid the additional source.

The extracted spectra were rebinned to 2 \AA~bin$^{-1}$ prior to polarimetry analysis. Then very fine wavelength alignment (to within $0.1$ \AA) was done using the cross-correlation of all eight available extracted spectra for a single object. The zero point of the polarization angle was calibrated by normalizing to the published value for HD15528 \citep{Clemens1990} of 105.13$^{\circ}$.  We find that calculated polarization values and angles for our polarized standards agreed well with published values. For HD15528 we measure a polarization of $4.979\%$ in the $V$-band (mean $\lambda \sim 5388$ \AA~and $\Delta \lambda \sim 352$ \AA) compared to $4.849\%$ for the originally published value \citep{Clemens1990}.  For HD204827, we measure a polarization of $5.760\%$ with a polarization position angle of $\theta = 59.48^\circ$ in the $B$-band (4000-5000 \AA). This is extremely close to the published values of $P = 5.648 \pm 0.022\%$ and $\theta = 58.20 \pm 0.11$ degrees in \citet{Schmidt1992a}. In addition, observations of unpolarized standard stars show well-behaved, flat response over the entire observed wavelength range and give a calculated polarization of 0.08\%. Though this measurement includes some combination of the stellar and instrumental polarization, it gives us an approximate upper limit on the effects of instrumental polarization.

We conducted a variety of consistency checks with our data to convince us of the accuracy of our results. We measured the polarization for each sequence of exposures separately to make sure they were consistent before we combined them using a simple algorithm which weights each half equally. We also checked that the normal and optimal weighting algorithms \citep{Horne1986} for the source extraction give similar results. We used the optimal weighting algorithm except when we measured the line polarization for SDSS1515+1757T and SDSS1623+3122TE where the results did not agree well. It is also encouraging that clear detections of polarization in the continuum showed consistent values of the observed polarization angle in adjacent wavelength bins except in cases where the polarization angle can clearly be seen to physically vary as a function of wavelength (see section \ref{sec:analysis} for more information).

All polarization measurements were made in the $q$ and $u$ spectra and converted to a debiased polarization and position angle, following \citet{Miller1988}.  To increase S/N in the continuum we binned over several hundred \AA ngstroms for each measurement. We were careful not to include wavelengths on the very red edge of the blue CCD to avoid the second order light due to the blue side grism.  Polarization fractions and angles as well as the wavelength range over which each was calculated are listed in Table \ref{tab:pol}. The usual definition of polarization is a positive-definite value, $p=\sqrt{q^2+u^2}$ which becomes problematic in our regime of low S/N.  Instead we report the debiased polarization which has a better behaved error distribution, $p= \pm \sqrt{\left |q^2+u^2-\sigma_q^2-\sigma_q^2\right |}$ where the overall sign is set by the value within the absolute value signs. Thus, if a measurement is dominated by measurement error, the quantity $q^2+u^2-\sigma_q^2-\sigma_u^2$ will have a value less than zero, and the debiased polarization will be negative. In cases where the debiased polarization is dominated by measurement error, we instead report 2$\sigma$ upper limits.  The total flux spectra as well as continuum polarization, q and u normalized Stokes parameters and position angle measurements are shown in figure \ref{fig:pol} as well as the Appendix.  
  
All our targets lie at relatively high Galactic latitudes with a range of $34.0^{\circ} < |b| < 71.5^{\circ}$, and thus the contribution to the polarization from interstellar dust is expected to be quite low. In fact, \citet{Serkowski1975} demonstrated that the Galactic interstellar polarization is $\lesssim 9\%\times$ $E$(B-V). We check the measured values of E(B-V) along the sightlines to our targets from both \citet{Schlegel1998} and \citet{Schlafly2011} and find values in the range 0.0206 $<E$(B-V) $<$ 0.1179 mag, implying interstellar polarizations of $\lesssim 1\%$, much smaller than any of the values we are measuring in our targets. Thus we conclude that interstellar polarization has a negligible effect on our measured polarization values.

\section{Results}
\label{sec:analysis}

Our sample spans a range of rest-frame optical properties as described in Section \ref{ssec:optical}, from classical Type 2 spectra with quiescent [\ion{O}{3}] kinematics to extreme [\ion{O}{3}] outflow activity.  Here, we discuss the results of our spectropolarimetry analysis and connect them to other known multi-wavelength properties of our targets.

\begin{figure*}
\includegraphics[scale=0.85,trim= 12 10 20 30,clip=true]{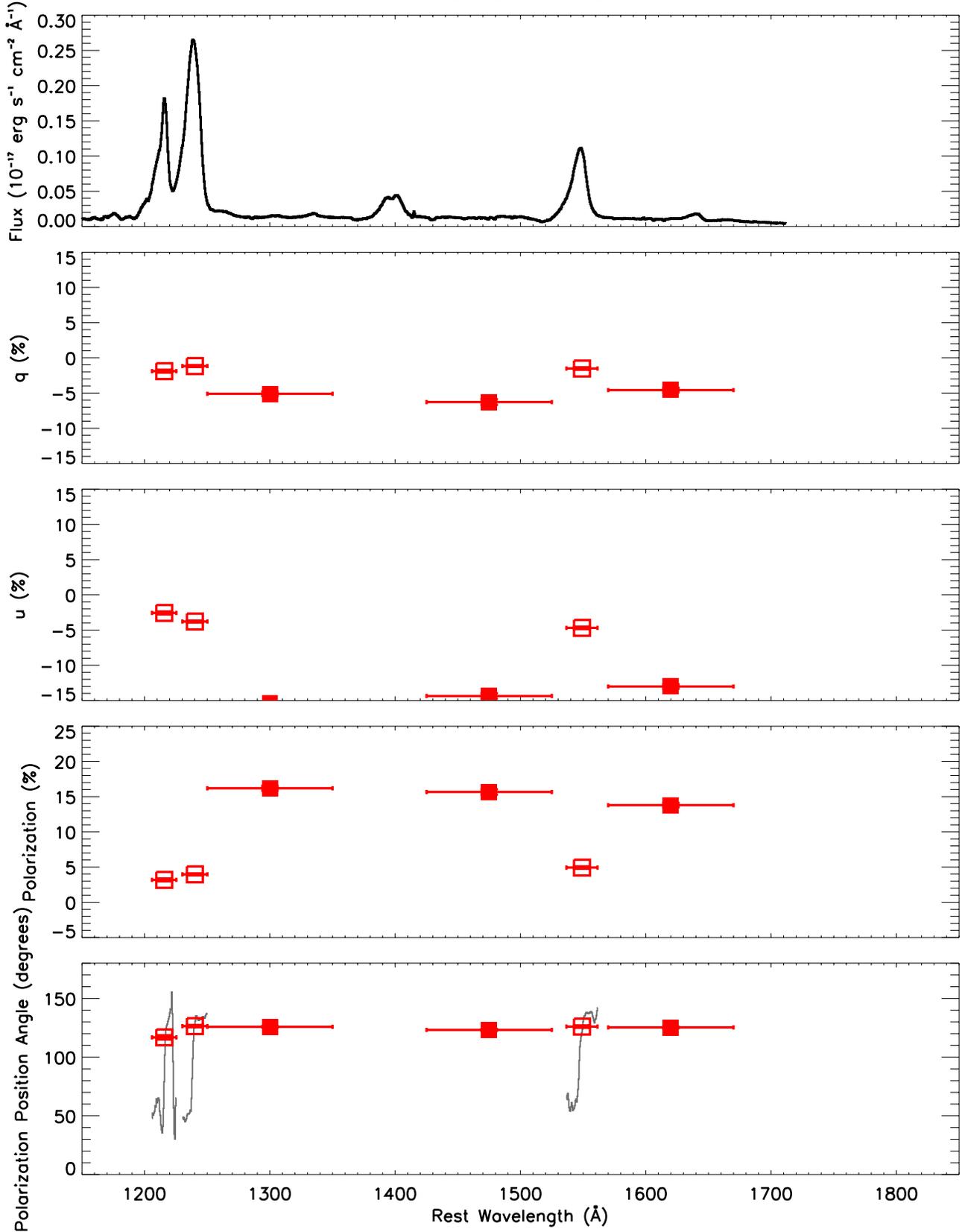}
\caption{\small LRISp spectra of our highest S/N target (all others are displayed in a similar manner in Appendix \ref{sec:app}). The top panel shows the total flux spectrum. The second and third panels show the normalized Stokes parameters $q$ and $u$ binned over the wavelength range as indicated by the horizontal axis error bars (same as Table \ref{tab:pol}).  The fourth panel shows the continuum polarization and statistical error in that measurement over the same bins, and the bottom panel shows the same for polarization angle. For data points where the debiased polarization is uncertain we plot the 2$\sigma$ upper limit. The filled squares represent continuum measurements while the open squares represent emission-line measurements. In grey, we include the smoothed polarization position angle (7-pixel bins) in the emission lines.}
\label{fig:pol}
\end{figure*}

In Figure \ref{fig:pol} and Appendix \ref{sec:app} we present the Keck
LRISp data for each of our objects. One object, SDSS16521+1728E
(Figure \ref{fig:pol}), has considerably higher S/N than the other
objects and thus we use it in what follows as an example of several trends. We display the high sensitivity total UV spectrum in the first panel. The following panels show binned quantities for the normalized Stokes parameters $q$ and $u$, the polarization fraction and polarization position angle in bins designated by the horizontal axis error bars (and listed in Table \ref{tab:pol}). We also display the full polarization position angle as a function of wavelength in the emission lines (smoothed using a least-squares polynomial smoothing filter with a window of seven pixels) to display variations within the lines as a function of velocity. In the following sections we discuss the trends we see in our spectra.  

\subsection{Level of Continuum Polarization}
\label{ssec:cont_pol}

In Table \ref{tab:pol}, Figure \ref{fig:pol} and the Appendix, we present continuum polarization measurements in several continuum regions, chosen in the rest-frame of the quasar to avoid contamination by UV emission lines. The level of continuum polarization just blueward of \ion{C}{4} (rest frame 1425--1525 \AA) spans from $15.7\pm0.4\%$ in SDSSJ1652+1728E to consistent with zero in SDSSJ2215$-$0056E. 

In our sample of five sources, four (both Type 2 quasars and two of the ERQs) are consistent with polarization fractions significantly greater than $3\%$ -- a high polarization fraction for non-synchrotron sources. Three of the five have polarization fractions $> 10\%$. Overall, the polarization values measured are much higher than those observed in Type 1 AGN. For example, \citet{Berriman1990} found an average polarization of $0.5\%$ for the 114 quasars from the Palomar-Green Bright Quasar Survey with a maximum polarization of $2.5\%$. This is expected as Type 1 AGN are dominated by direct quasar light that would dilute any polarization signal. In \citet{Alexandroff2013}, we measured the polarization of two Type 2 quasar candidates but unlike our current sample designated with a `T', these objects were not pre-selected to be unambiguous Type 2 quasars based on their rest-frame optical spectra.  We found polarization $> 5\%$ in the blue continuum of one object, SDSSJ220126.11+001231.5, and an average polarization further redward of $1.9 \pm 0.3 \%$, higher than could be accounted for by a typical unobscured quasar combined with instrumental systematics ($\lesssim 0.1 \%$) and Galactic polarization ($< 0.8 \%$). The second object observed in that paper, SDSSJ004728.77+004020.3 had an average continuum polarization of only $0.91 \pm 0.35\%$, making our interpretation of the nature of the source less clear. Contamination from unpolarized quasar continuum is possible in these objects.  

Perhaps the best context for our sample is a direct comparison to
other verified Type 2 AGN, radio galaxies, or other samples of red
quasars.  \citet{Zakamska2005} measure a polarization $> 3\%$ in nine
of twelve luminous Type 2 quasars at $z\sim 0.5$ and a mean
polarization of 5.7\% for all twelve objects, with a maximum
polarization of $15.4\%$ at rest wavelengths between 2800\AA\ and
6500\AA. They contend that the high polarization is due to dust
scattering and that all but one object are essentially uncontaminated
by the host galaxy (which is thought to be the main dilution mechanism
of the intrinsic polarization fraction in Type 2 AGN).  

Two IR-bright Type 1 quasars, IRAS 13349+2438 and IRAS 14026+4341
(also a BAL quasar), have been studied with polarimetry in the
rest-frame UV \citep{Hines2001}. They exhibit high polarization
fractions at their bluest wavelengths (9.44\% between rest-frame
2000-2700 \AA~and 14.84\% between rest-frame 1690-2270
\AA~respectively) while reaching a maximum polarization at around 3000
\AA~with the host galaxy contributing at most 10\% of the light at
these wavelengths. Fifty percent of the entire IR-bright sample
observed in optical polarization show polarization fractions $> 3\%$. Similarly, \citet{Smith2000,Smith2002, Smith2003} studied the polarization properties of a sample of two Micron All-Sky Survey (2MASS)-selected red AGN with $M_{K_s} < -23$ and $J-K_s > 2$. This sample includes both Type 1 and Type 2 AGN with a mean redshift of $z = 0.25$. \citet{Smith2002} showed that in a sample of 70 2MASS quasars with optical broadband polarimetry, 10\% were polarized with $p > 3\%$ with a maximum polarization measured of 11\%. At these redshifts, host galaxy light is likely a significant contaminant to the continuum luminosity ($\gtrsim 50\%$), so the intrinsic polarization fractions are likely higher. 

We argued in \citet{Alexandroff2013} that the rest-frame UV continuum in high-redshift Type 2 candidates from SDSS was too luminous to be explained by the host galaxy alone and too blue to be explained by the reddening of the intrinsic quasar light. There were two competing hypotheses for the origin of this emission. The first is that the observed UV continuum may be due to a ``patchy" geometry allowing some direct sightlines to the quasar. The second is that the central quasar is completely obscured from view, while all of the UV continuum is due to scattered light. Similar ideas were discussed for the origin of the rest-frame UV continuum of ERQs and other red quasars \citep{Assef2016, Zakamska2016, Hamann2017}. From the comparison with other available samples we find that the continuum polarization of our targets is at the high end of the polarization fraction distribution for obscured and red quasars. Detecting such high values of continuum polarization in our targets is unambiguous confirmation that most of the continuum emission is due to anisotropic scattered light. 

Another paradigm we can test with our data is that of the geometry of scattering. Is the circumnuclear obscuration axisymmetric, in which case we expect an elongated scattering region with relatively minor geometric cancellation of polarization and therefore high polarization values? Or is the nucleus largely enshrouded, with several randomly oriented scattering regions \citep{Schmidt2007}, where we can expect geometric cancellation of polarization and therefore low polarization values? With our admittedly small sample, we find no relationship between the rest-frame optical type (Type 2, ERQ) or the presence of [\ion{O}{3}] outflows, and the continuum polarization level which might be expected if a difference in optical type or outflow velocity denoted a difference in quasar evolutionary state or obscuring geometry.

\begin{table*}[htp]
\caption{Continuum and line polarization fractions \label{tab:pol}}
\begin{center}
\begin{tabular}{lrrrrrrrr}
\hline
Short Name & 1250-1350\AA & 1425-1525\AA & 1550-1650\AA & 1650-1750\AA & 1750-1850\AA & Ly$\alpha$ & \ion{N}{5} & \ion{C}{4} \\
\hline
SDSSJ1232+0912E &7.7$\pm$1.9&12.1$\pm$4.0&12.4$\pm$2.0&17.4$\pm$2.7&17.0$\pm$1.5&7.5$\pm$0.8&3.4$\pm$0.4&5.8$\pm$0.4\\
SDSSJ1515+1757T &$<$6.5&11.0$\pm$3.2&$<$1.8&9.4$\pm$2.6&10.2$\pm$3.5&8.8$\pm$0.9&5.3$\pm$3.5&7.0$\pm$1.3\\
SDSSJ1623+3122TE &$<$5.2&6.5$\pm$3.8&8.5$\pm$1.7&7.0$\pm$3.0&2.1$\pm$2.3&9.6$\pm$1.0&5.3$\pm$3.0&5.6$\pm$1.1\\
SDSSJ1652+1728E &16.2$\pm$0.3&15.7$\pm$0.4&13.8$\pm$0.3&&&3.2$\pm$0.2&4.0$\pm$0.1&4.9$\pm$0.2\\
SDSSJ2215$-$0056E &$<$14.7&$<$4.9&$<$1.3&$<$6.9&$<$9.1&1.8$\pm$1.3&1.5$\pm$2.0&$<$1.2\\
\hline
\end{tabular}
\end{center}
\begin{tablenotes}
\item[a] Polarization fractions of our targets as a function of rest wavelength. All measurements are listed as percentage of the total flux. For emission lines the polarization was measured over the wavelength range corresponding to the FWHM of \ion{C}{4} for each object. Upper limits are 2$\sigma$.
\end{tablenotes}
\end{table*}%

\subsection{Wavelength Dependence of Continuum Polarization}
\label{ssec:wave_pol}

Scattering of the continuum can be due to dust or free electrons in the ionized gas. The expected polarization fraction due to electron scattering is described by the Thomson formula and is independent of wavelength. For dust scattering, the polarization fraction may vary as a function of wavelength depending on the dust composition and size distribution (e.g., \citealt{Zubko2000}). In this section we discuss the trends seen in our sample, with more details discussed in section \ref{ssec:res_scattering}.

There is marginal evidence for a wavelength dependence in the
polarization fraction in the continuum of our objects, with the
polarization fraction appearing to rise at longer wavelengths. Such a
trend is particularly evident in SDSSJ1232+0912E.  Unfortunately, the
limited S/N of our data makes it difficult to determine if this trend is real, as most of the variation is within the observed error bars.  \citet{Alexandroff2013} found that there was evidence for wavelength-dependent polarization fraction in one object that was strongest at bluer wavelengths, the opposite of what we observe in SDSSJ1232+0912E.  

An increasing polarization fraction at redder wavelengths suggests either that the polarization is wavelength-dependent and therefore likely caused by dust, or that the continuum is contaminated by blue unpolarized light (such as unpolarized quasar continuum) that dilutes our polarization signal at bluer wavelengths.  However, the evidence for this wavelength dependence is marginal at best so we will defer any discussion of wavelength-dependent polarization in our objects for future work.

\subsection{Overall Polarization of Emission Lines}
\label{ssec:line_pol}

Photons with frequencies within the emission lines can be scattered by electrons and by dust, just like continuum photons, but also resonantly, if they are incident on ions with velocities such that their Doppler-shifted transition frequencies are an exact match to the incident frequencies. We first consider the possibility that the dominant mechanism of polarization of emission lines is the same as that of the continuum polarization (i.e., likely dust scattering). 

In this case, the observed level of polarization in emission lines relative to the continuum is a good indicator of the geometry and polarization mechanism of the emission lines. In particular, the polarized fraction of the emission lines relative to the continuum can indicate the scale of the scatterer. If the emitting region and the scattering region are on the same scale, the polarization is diluted by geometric effects as there is no single scattering direction. Higher values of the ratio of emission line to continuum polarization imply scattering on large scales, beyond the line emitting regions, so that the lines remain as polarized as the continuum. Similarly, lower values of this ratio imply that the scattering is on scales closer to or within the emission-line region and thus the line emission is only marginally polarized.  Geometric dilution in emission-line regions relative to the continuum is found in many AGN sources \citep[e.g.][]{Stockman1981,Glenn1994,Goodrich1995b,Tran1995c} and implies that the scattering region is on a similar, or slightly larger scale, than the emission region.

We consider the UV emission lines of Ly$\alpha$, \ion{N}{5} and
\ion{C}{4}, integrating over the FWHM of \ion{C}{4} (see Table
\ref{tab:info}).  The observed polarization fractions of the emission
lines are lower than the continuum polarization for four of our
objects, the exception is SDSSJ1623+3122TE. Nevertheless, polarization values can still reach levels of $> 8 \%$ in the Ly $\alpha$ emission line, most notably in SDSSJ1515+1728T and SDSSJ1623+3122TE. 

We calculate the ratio of the line polarization fraction (for Ly$\alpha$, \ion{N}{5} and \ion{C}{4}) to the continuum polarization fraction just blueward of \ion{C}{4} between 1425 \AA~and 1525 \AA~and find some evidence that those objects designated as classical Type 2 quasars in our sample show higher values of this ratio. The mean and the sample standard deviation over all six lines for our two ERQs (excluding SDSSJ2215$-0056$E where there is not a high enough S/N ratio in the lines) is $0.36 \pm 0.14$, whereas for the two Type 2 quasars it is $0.85\pm 0.31$. The ratio of the line to continuum polarization measures the geometric mixing of polarization as emission propagates from the emission-line region and scatters off the scattering medium. Thus, our measured ratios imply that the scattering regions for classical Type 2s may be on larger scales than the scattering regions in ERQs. 

This geometric interpretation of line polarization dilution becomes more complicated if resonant scattering has a strong or dominant contribution. Photons that are resonantly scattered just once and then leave the scattering cloud have the same polarization efficiency and phase function as Thompson scattering, i.e., these photons can be expected to be highly polarized. Those photons that scatter multiple times are essentially unpolarized, so they also dilute the net polarization \citep{Lee1994b}. We further discuss emission-line scattering processes in section \S~\ref{ssec:res_scattering}. 

\subsection{Kinematics of Line Polarization}
\label{ssec:kine_pol}

\begin{figure*}
\includegraphics[scale=0.7,trim=32 70 10 120,clip=true]{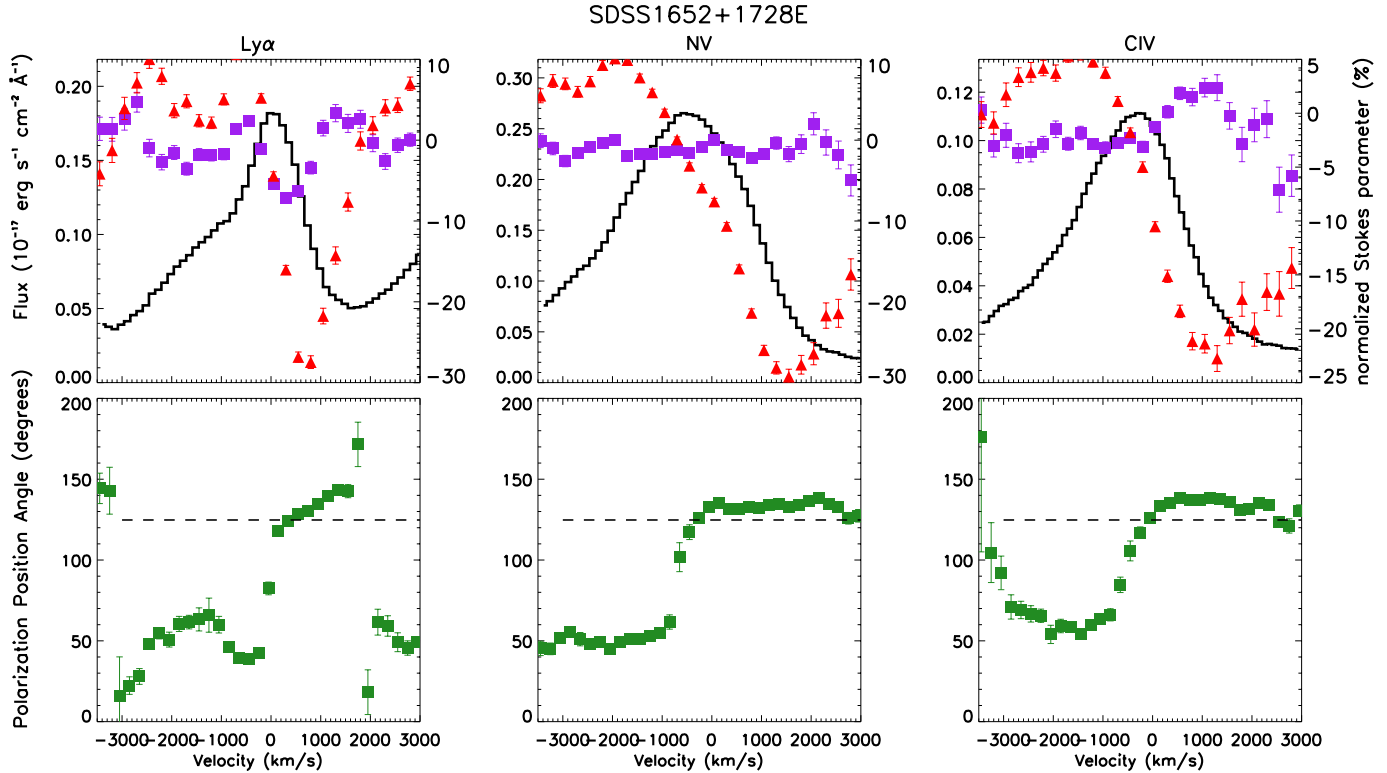}
\caption{\small LRISp spectrum of SDSSJ1652+1728E over the emission
  lines of Ly$\alpha$, \ion{C}{4} and \ion{N}{5} in velocity space
  ($\pm 3000$km s$^{-1}$). A similar figure for the remaining four
  objects can be found in the Appendix, Figures
  \ref{fig:velocity_app1}-\ref{fig:velocity_app2}.  We plot the total
  flux of the line (black histogram) as well as the normalized Stokes
  parameters $q$ and $u$ in percent (purple square and red triangular
  points) in bins of 200 km s$^{-1}$ as well as the polarization
  position angle (green square points) in similar bins below. Finally,
  the black dashed line shows average polarization position angle from the continuum bins.}
\label{fig:velocity}
\end{figure*}

\begin{figure*}
\includegraphics[scale=0.85,trim=20 30 10 10,clip=true]{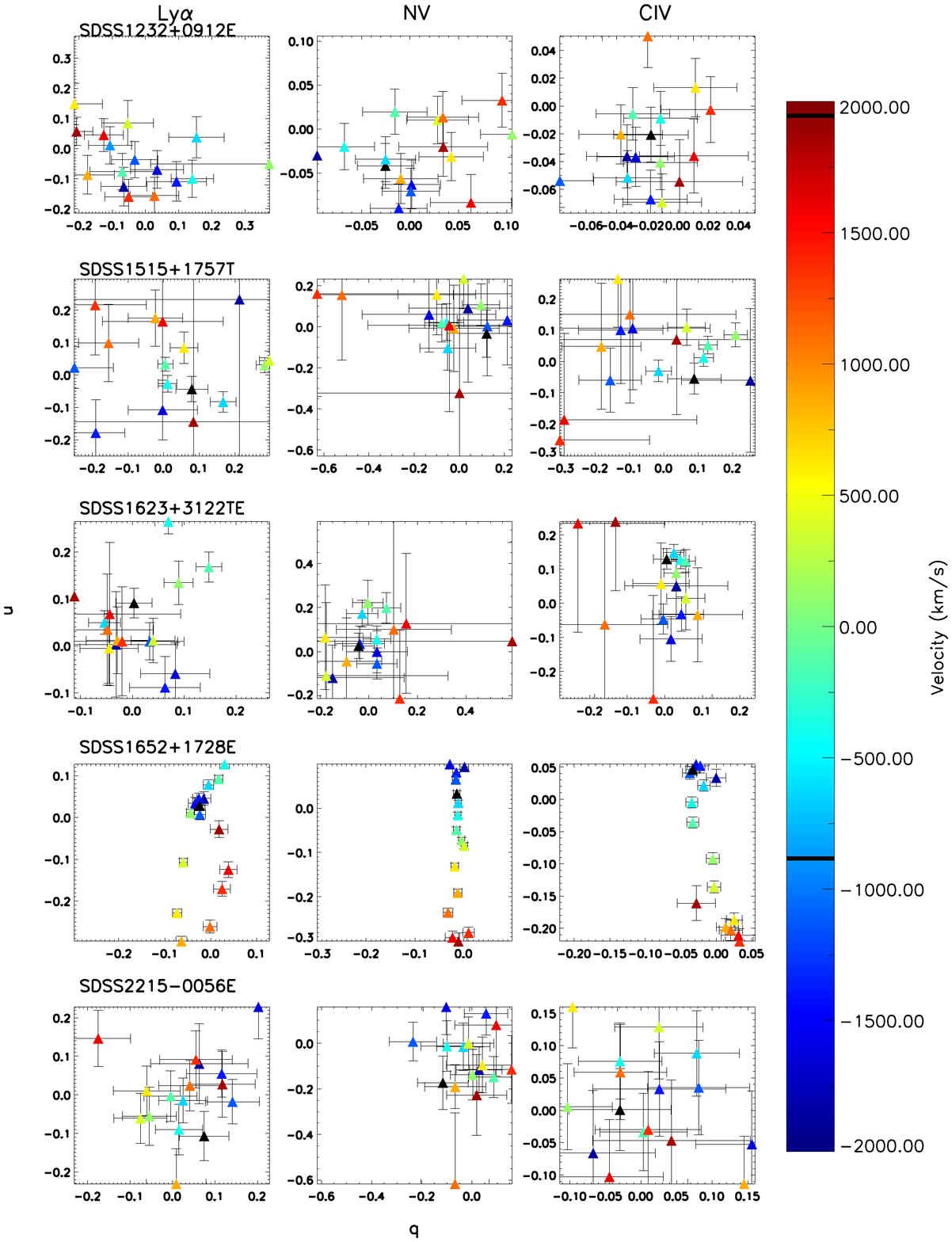}
\caption{\small Normalized Stokes parameters $q$ and $u$ for each of
  the key emission lines in our targets.  Each point represents a 250
  km s$^{-1}$ bin. We show over $\pm 2000$ km s$^{-1}$ in velocity
  space with the colour scale. In SDSSJ1652+1728E, there is a loop in
  $q$ vs. $u$ space, most apparent in Ly$\alpha$. For the other objects a potential loop can be identified by a gradient in the colour (velocity) across the $q-u$ plane.}
\label{fig:q_u}
\end{figure*}

We observe a significant variation in the polarization fraction across emission lines. The polarization fraction tends to reach a minimum around the line centroid and a maximum at a redshift between +1000 and +2000 km s$^{-1}$. This pattern is perhaps most pronounced in SDSSJ1652+1728E (Fig. \ref{fig:velocity}). Such wavelength-dependent polarization in emission lines is present in several objects in our sample and may provide a unique window into the scattering geometry.  We discuss this further in section \S~\ref{sec:discussion}.

Perhaps the most intriguing trend we observe is the wavelength
dependence of the polarization position angle across all emission
lines. This trend can be observed in SDSSJ1515+1728T, SDSSJ1623+3122T
and most dramatically in SDSSJ1652+1728E. This observed trend has
several key features. First, the polarization position angle appears
to vary by almost exactly 90 degrees across the emission line
(see Fig. \ref{fig:velocity}, Fig. \ref{fig:velocity_app1} and
Fig. \ref{fig:velocity_app2}).  Most notably for our analysis, the
polarization position angle of the continuum is the same as the
polarization position angle of the emission lines at their red
wings. This trend can easily be observed in Figure \ref{fig:velocity},
but is also apparent as a change in the ratio of the $q$ and $u$
normalized Stokes parameters in Figure \ref{fig:q_u}. This format is
inspired by similar figures in papers describing polarization in supernovae \citep[SNe; for a review see][]{Wang2008}. 

If there were a single, smooth and axisymmetric scattering region where the polarization was independent of wavelength, we would find a polarization fraction independent of wavelength, which would result in a clustering of all data points in the $q-u$ plane. If the axisymmetric scatterer (which may have a poloidal velocity distribution) has a wavelength-dependent optical depth, then at each wavelength the viewer would see a different slice or shell in the axisymmetric scattering material. Therefore, the degree of polarization may vary from one shell to the next (or from one wavelength to the next), while the axis of symmetry remains constant. As a result, the wavelength-dependent locus in the $q-u$ plane is a straight line, also known as the ``dominant axis'' \citep{Hoeflich1996, Wang2001}. If the line goes across the (0,0) point in the $q-u$ plane, this implies a polarization position angle change of 90 degrees. An excellent example of this behavior is the \ion{N}{5} emission line in SDSSJ1652+1728 (Figure \ref{fig:q_u}). Despite the much larger error bars, we can see a similar trend in the \ion{N}{5} emission line in SDSSJ1232+0912, where $q$ and $u$ are negative on the blue wing of the line and positive on the red one.  

Deviations from the dominant axis in the $q-u$ plane then represent deviations from the assumptions of smoothness or axisymmetry in the scattering geometry as seen most prominently in the Ly$\alpha$ and \ion{C}{4} emission lines of SSSJ1652+1728. These deviations must be finite to maintain the presence of a single dominant axis in the plane (e.g., the presence of clumpy material). Such loops are also seen in the polarization of supernovae. Possible explanations for loops in the $q-u$ plane include an overall asymmetry to the scattering structure, an additional expanding shell with a different geometry, or the breakup of an axially-symmetric scattering structure into clumps \citep{Kasen2003, Wang2008}. 

Understanding the origin of the variations in polarization and its position angle within the AGN emission lines is of great interest because these variations may allow us to probe the geometry of AGN on scales no accessible to other observations. Such changes have been widely observed in AGN, though often the polarization angle varies only by a few to a few tens of degrees, whereas the objects presented here -- most notably SDSSJ1652+1728 -- show an extreme 90 degree polarization angle swing. \citet{Smith1995} argued that structure in the polarization fraction of the emission lines in Mrk 231 requires several scattering components.  \citet{Smith2000} further argued that structure in the polarization fraction and polarization position angle across broad emission lines, coupled with larger values of polarization in the continuum than the emission lines, implies that the BLR of the 2MASS quasars contains multiple scattering components.  In contrast, \citet{Young2000}, \citet{Smith2005}, and \citet{Young2007} argued that such swings in the polarization position angle were due to emission and scattering by an equatorial disk wind that is visible in Type 1 quasars but is overwhelmed by polar scattering signatures in Type 2 quasars. 

The extreme (close to ninety degrees) rotation in the polarization position angle in our objects as a function of velocity within the emission lines implies dramatic changes in the scattering geometry as a function of velocity. The linear features in the $q-u$ plane may be successfully explained by axisymmetric models, which we discuss in the next section. In contrast, loops in the $q-u$ plane imply deviations from axisymmetry, which we do not attempt to model here. 

\section{Discussion}
\label{sec:discussion}

Several trends are visible in our observations as noted in section \ref{sec:analysis}:
\begin{enumerate}
\item polarization reaching $> 10\%$ in the continuum for three of five sources;
\item a ratio of emission-line polarization to continuum polarization that is $< 1$ for all sources, especially ERQs;
\item a change in the polarization position angle across emission lines of $\sim$90 degrees, with the redshifted line emission at the same position angle as the continuum.
\end{enumerate}

\subsection{Proposed Model}
\label{sec:model}

To model these trends, we consider models with a number of ingredients. Polarimetry models include both an emission and scattering region (though they can be coincident) which can be either stationary or have some velocity. In addition, we can vary the type of scatterer and scattering mechanism. For example, the classical model that explains the spectropolarimetry of obscured quasars is that of a point-like emitting region with a conical scattering region defined by the opening angle of the torus \citep{Miller1991}.

\begin{figure*}
\includegraphics[scale=0.3]{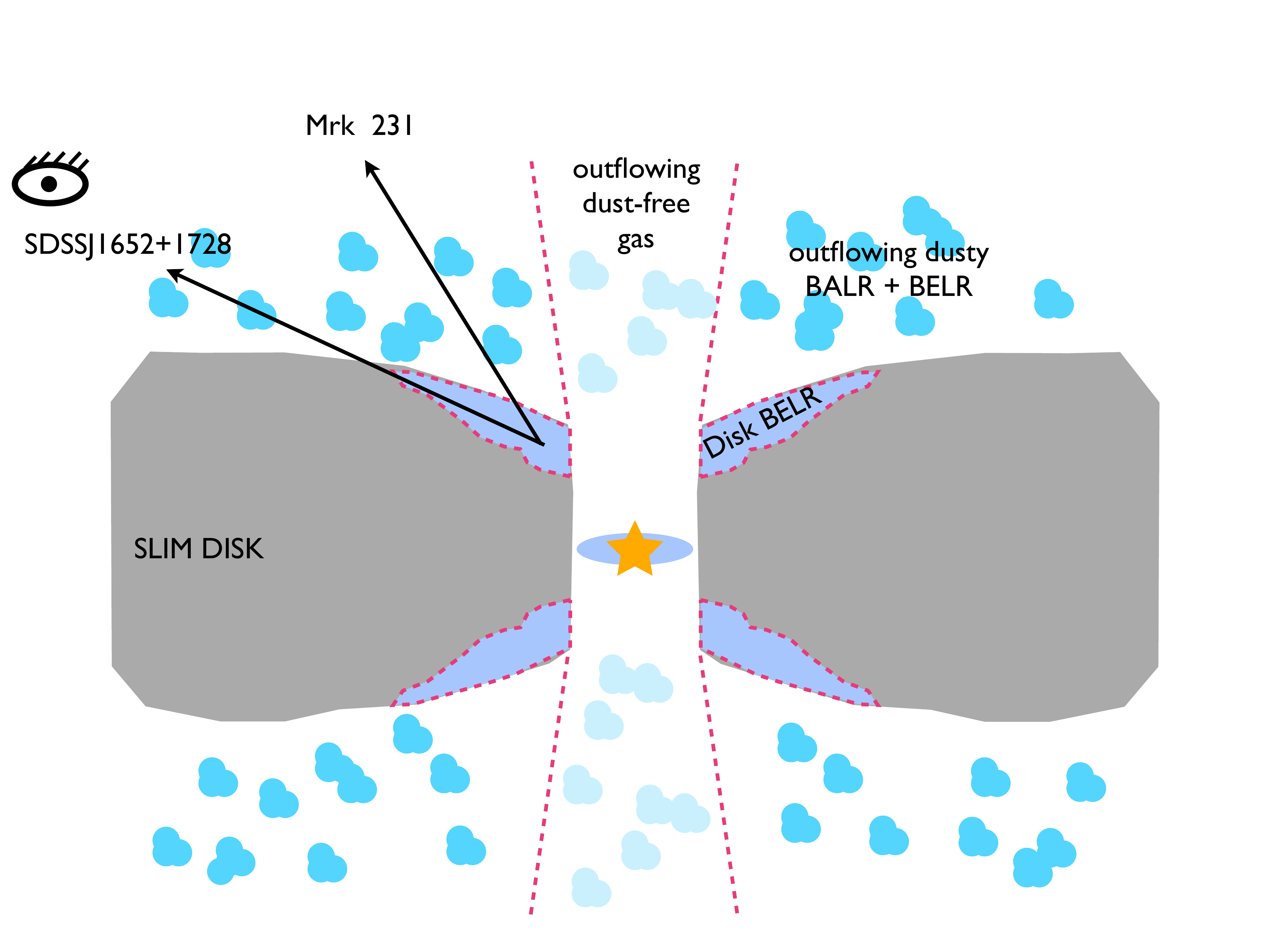}
\includegraphics[scale=0.3]{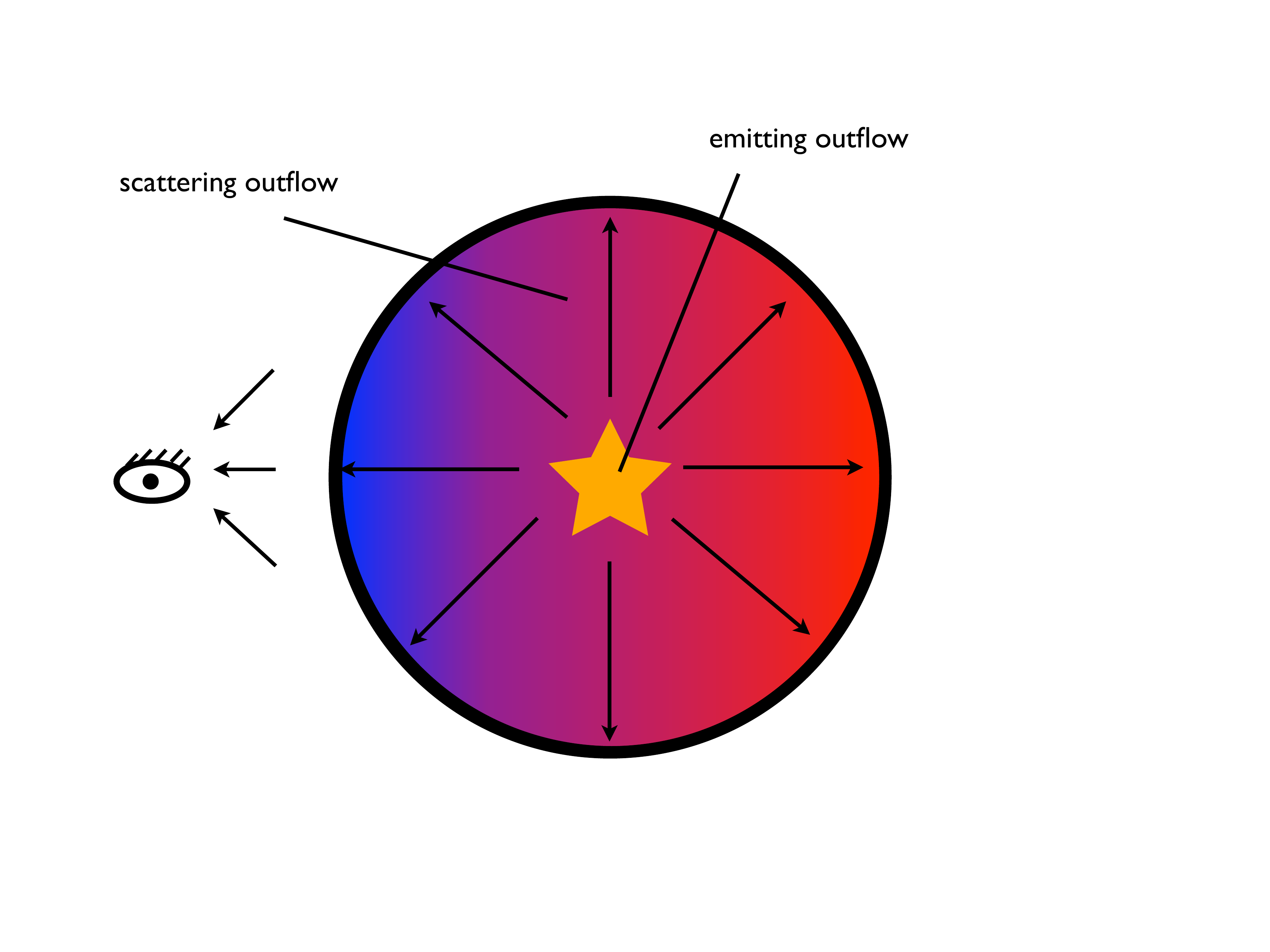}
\caption{\small Our preferred model for SDSSJ1652+1728 adapted from \citep{Veilleux2016}. It combines a polar outflow which produces
  the emission lines with a dusty equatorial wind where the scattering
  occurs. \textbf{Left:} Side-on view of the inner $\sim$10 pc of our
  model (not to scale). The inner accretion disk ($\lesssim 0.01$pc)
  is responsible for the X-ray emission and is surrounded by a
  geometrically-thick accretion disk (``slim disk") which, for a near-
  or super-Eddington quasar produces narrow funnels that drive a polar
  wind. This polar wind is comprised of both dust-free outflowing gas,
  responsible for X-ray absorption, as well as dusty clouds that
  produce FeLoBAL absorption signatures and the blueshifted FUV
  emission lines of Ly$\alpha$ and \ion{C}{4}. The extended accretion
  disk is the source of the UV and optical continuum emission as well
  as the Balmer lines (seen through the screen of the outflowing dusty
  BALR).  Mrk 231 is constrained to have a jet angle of
  $\theta_{max}=25.6 ^{+3.2}_{-2.2}$ degrees from radio observations \citep{Reynolds2013}. This means that the scattered light, though
  still significant, is unpolarized in the observer's line of sight
  due to geometric cancellation.  In contrast, SDSSJ1652+1728 is seen
  at a larger angle to the jet or polar axis as inferred from the mean
  SED of ERQs (see section \ref{sec:model} for further
  discussion). The increased column density of obscuration suppresses
  the FeLoBAL features while the more edge-on sight line reduces
  geometric cancellation of the polarization signature. Though the
  original cartoon from \citet{Veilleux2016} contains a radio jet we
  omit it as irrelevant to this paper.  \textbf{Right:} The same model as viewed from above (looking along the polar axis). Here we show how the kinematic features in the polarized emission-line regions are created by the blue- and red-shifted scattering emission respectively. The viewer (eye on the left) sees blueshifted emission coming towards them and scattered off of clouds moving towards the observer. In contrast, the redshifted wings of the emission lines are dominated by scattering off the ``sides" of the outflow (from the observer's perspective). Emission from the back of the cone is largely suppressed in polarization due to intervening extinction. These features combined produce the ninety degree rotation in the polarization position angle across the emission lines.}
\label{fig:pol_model}
\end{figure*}

Combining all of these considerations, we constructed a model of the emission, scattering and resulting polarization expected for simple geometries and polarization mechanisms.  We find that the model with the best fit to our data is a polar emission region (expanding emission within a filled cone) which is scattered by dust or resonantly in an equatorial outflow (see Figure \ref{fig:pol_model} and \citealt{Veilleux2016}). The full details of this model will be described in Zakamska et al 2018 (in prep). While there are certainly more complicated geometries that would reproduce all our results, this model has the advantage of being straightforward and physically motivated.

There is both observational evidence and extensive theoretical work to support an equatorial, dusty outflow from the AGN torus that is supported by radiation pressure \citep[e.g. ][]{Wills1992,Elitzur2006,Veilleux2016,Chan2016,Elvis2017}.  If the AGN is near or super-Eddington, the creation of a geometrically thick accretion flow also produces polar radiation driven outflows \citep[e.g.][]{Abramowicz1988,Scadowski2015,McKinney2015}. We adapt the cartoon of \citet{Veilleux2016} who model Mrk 231 with a slim disk responsible for the continuum emission. The broad emission-line region (BELR) is composed of a polar region, an extended accretion disk atmosphere and the disk itself \citep[see][figure 8 and our Figure \ref{fig:pol_model}]{Veilleux2016} whereas the broad absorption line region (BALR) originates only in the polar outflow and disk atmosphere.  Using this model, the FeLoBAL features are produced in the BALR, blueshifted FUV emission lines are produced in the outflowing BELR, and the optical and NUV emission lines are produced in the disk BELR.  We find that some of the features of this model reproduce our spectropolarimetry results well. In particular, when the FUV lines are produced in the polar region and then are scattered by a dusty, ionized equatorial outflow, the puzzling kinematics of the polarized emission lines can be reproduced. 

The main difference between Mrk 231 and our objects (ERQs and Type 2s) is that our viewing angle is closer to edge-on than the $\sim$25$^\circ$ polar viewing angle inferred for Mrk 231 from radio observations \citep{Reynolds2013}. We make this conclusion on the basis of the shape of the spectral energy distributions (SEDs) of our sources \citep{Zakamska2016,Hamann2017}: as a function of wavelength, ERQ SEDs do not rise until 1-2$\mu$m, whereas Mrk 231 is essentially unobscured at red wavelengths in the optical, and the steep rise of the Mrk 231 SED occurs at $\sim 3000$ \AA. Therefore, the net line-of-sight column density of obscuration is $\sim 10$ times higher in ERQs than in Mrk 231, implying the more edge-on viewing angle. For a Small Magellanic Cloud (SMC) dust curve \citep{Weingartner2001,Draine2003} in the ``screen of cold dust'' approximation, the Mrk 231 rise would correspond to $A_{\rm V}\simeq 0.3$ mag, and the 1$-$2$\mu$m rise for ERQs would correspond to $A_{\rm V}\simeq 5$ mag. 

The difference in viewing angle could explain why Mrk 231 exhibits low levels of FUV polarization.  At the viewing angle of Mrk 231 the scattering region becomes nearly symmetric and thus geometric cancellation would dilute the polarization fraction (similar polarization fractions with opposite scattering angles will combine to produce a lower total polarization fraction) though the scattered flux is still high. In contrast, for the nearly edge-on objects in our sample the equatorial scattering region is extended in the plane of the sky.  

In subsequent sections we explore how this model matches the observational trends noted in our data. Section \S~\ref{ssec:disc_contpol} explains how this model matches our continuum polarization results. In Section \S~\ref{ssec:res_scattering} we discuss the relative importance of resonant and dust scattering for the emission linesm and in Section \S~\ref{ssec:disc_linepol} we discuss the observed line polarization in the context of our geometric model. 

\subsection{Continuum Polarization and Scattering Geometry}
\label{ssec:disc_contpol}

While scattering off of dust or electrons is the most likely source of polarization in AGN, especially at the high levels we observe in our objects, without some understanding from imaging of the scales on which polarization occurs it is difficult to distinguish between the two. For objects where the scattering cone is visible on $\sim$kpc scales \citep{Dey1996,Kishimoto2001,Zakamska2005} the dominant scatterer is likely dust. In these cases, the required gas mass for electron scattering is likely prohibitively high though it is not always possible to completely rule out electron scattering. In contrast, on circumnuclear scales the polarization of NGC 1068 is wavelength independent from the X-ray to the optical, and is therefore likely due to electron scattering \citep{Miller1991}.

The high infrared-to-optical ratios and multiple features of the rest-frame optical spectra of our objects \citep{Alexandroff2013,Ross2015,Zakamska2016,Hamann2017} indicate that they are obscured. We also know that the observed UV continuum of our objects is dominated by scattered light, because the net continuum polarization is so high that it cannot be appreciably diluted by the direct (unpolarized) light of the quasar. Therefore, scattering must be happening primarily on scales greater than obscuration scales, which in turn must be larger than the dust sublimation region \citep[defined by the radius, $r_{\rm dust}$;][equation 5]{Barvainis1987}:

\begin{equation}
r_{\rm dust}=1.3\left(\frac{L_{\rm UV}}{10^{46}{\rm erg/sec}}\right)^{1/2}\left(\frac{T}{1500{\rm K}}\right)^{-2.8}{\rm pc}.
\end{equation}
Here $L_{\rm UV}$ is the UV luminosity and $T$ is the dust sublimation temperature. In the case of our extremely luminous sources ($L_{\rm UV} \approx$ a few $\times 10^{47}$ erg/sec), the sublimation scales might reach a few to 10 pc. In ERQs, these scales are consistent with the scales of obscuration inferred from X-ray observations \citep{goul18}.

For a normal dust-to-gas ratio, the cross-section of dust scattering is two orders of magnitude higher than the electron scattering cross-section \citep{Draine2003}, therefore since scattering must be happening outside of the dust sublimation zone, the presence of dust implies that dust scattering dominates our scattered light. In our proposed model, therefore, the dust scattering occurs somewhere in the BELR/BALR region (see Figure \ref{fig:pol_model}).

\begin{figure}
\includegraphics[scale=0.5,trim=100 240 0 100,clip=true]{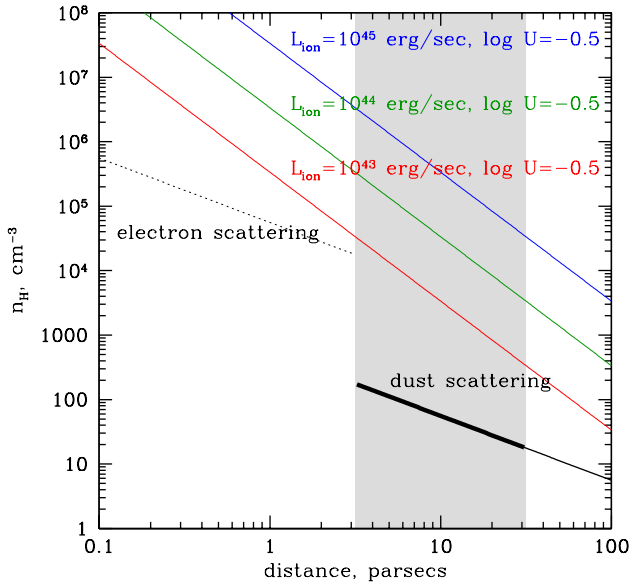}
\caption{\small The scattering efficiency and ionization parameter allow us to place constraints on the size of the scattering region and the density of scattering material.
Using equation (\ref{eq:epsilon}) and assuming a scattering efficiency
of 3\%, the black lines show the range of densities for reasonable
distances to the dust (solid) or electron (dotted) scattering regions
respectively. The coloured lines in the upper right show the limits on
the density placed at the same distances based on our assumptions of
the quasar ionizing luminosity and ionization parameter (equation
\ref{eq:ionization_parameter}). Our preferred regime is the grey
region which covers dust scattering at scales 3-30 pc (around the dust
sublimation radius). We are not able to place strong limits on the actual ionizing luminosity available from our quasars which is, additionally, likely suppressed by extinction. It is clear from this figure that the two limits, derived from the scattering efficiency and the ionization parameter, are incompatible. We hypothesize that this is because while the line emission occurs in dense clouds located in the quasar BELR, the scattering occurs on similar scales but in the less dense medium surrounding these clouds.}
\label{fig:dust_constraints}
\end{figure}

The scattering efficiency, the ratio of scattered flux to the flux that would have been observed directly in the absence of obscuration, is defined as 
\begin{equation}
\epsilon=\frac{{\rm d}\sigma}{{\rm d}\Omega}\Delta \Omega \int n_{\rm H}(r) {\rm d} r.
\label{eq:epsilon}
\end{equation}
Here ${\rm d}\sigma/{\rm d}\Omega$ is the cross-section of scattering
per unit hydrogen atom as calculated in \citet{Draine2003}, $\Delta
\Omega$ is the solid angle covered by the scatterer as seen from the
emitter, and $\int n_{\rm H}(r) {\rm d} r$ is the column density of
hydrogen associated with the scattering region. For a constant
velocity outflow, $n_{\rm H}(r)\propto 1/r^2$, and therefore this
integral is weighted toward the smallest unobscured sizes.  We thus
approximate it as $n_{\rm H, max} d_{\rm min}$. We use the ratio of
the observed UV emission of ERQs to that of a type 1 quasar SED
normalized in the IR \citep{Richards2006} as an estimate of the
scattering efficiency, yielding $\epsilon=3\%$ (figure 16 of \citealt{Zakamska2016} and figure 8 of \citealt{Hamann2017}).  We take $\Delta \Omega=4\pi/3$ and 90$^{\circ}$ scattering by SMC dust with ${\rm d}\sigma/{\rm d}\Omega\simeq 4\times 10^{-24}$ cm$^2$/H/sr at 1500 \AA~\citep{Draine2003}. We arrive at the constraint in the density-scale plane in Figure \ref{fig:dust_constraints}, implying hydrogen densities for scattering material on the order of $10-100$ cm$^{-3}$.

If this scattering region were too dusty, then no emission from the
quasar could escape at all. Therefore, we need to verify that our
derived densities and sizes are compatible with the escape of
radiation from the quasar in order to be available for scattering into
our line of sight. The optical depth to dust extinction through the
scattering region is given by:
\begin{equation}
\tau_{\rm ext}=C_{\rm ext}\int n_{\rm H}{\rm d}r=\frac{\epsilon C_{\rm ext}}{\Delta \Omega {\rm d}\sigma/{\rm d}\Omega},
\label{eq:scattering_efficiency}
\end{equation}
where $C_{\rm ext}=2.9\times 10^{-22}$ cm$^2$/H is the total
cross-section for extinction \citep{Weingartner2001}. With our
fiducial values, $\tau\simeq 0.5$ and is therefore just right: a
quasar viewed through the BELR / BALR would be reddened, as is Mrk
231, but not completely extinguished. We also see from this expression
that the scattering efficiency, $\epsilon$, cannot be much greater
than our fiducial value, because otherwise the extinction becomes
prohibitively high. Therefore we conclude that our observed scattered
component cannot be appreciably obscured by either diffuse or patchy
material. 

Assuming an SMC-like dust distribution, the theoretically achievable
maximum polarization fraction at 1800 \AA~is  $\sim 14.7\%$ for a
scattering angle of 90$^{\circ}$ and the polarization fraction
continues to rise towards the blue \citep{Draine2003}. Thus, some of
our targets display such high levels of polarization in the continuum
that they come close to reaching this theoretically achievable limit.
The geometric cancellation of polarization means that to achieve the
observed 15\% polarization level, the intrinsic polarization level of
each scattering must be even higher. We know dust can be an efficient
polarizer depending on the size distribution of the dust as compared
to the wavelength of the scattered light
\citep{Weingartner2001}. Thus, achieving the observed polarization
fraction in our sources might require some adjustments to the dust
size distribution compared to pre-existing models. Adjustments from
the \citet{Draine2003} models may also be necessary to explain any
increase in the polarization fraction of our objects towards redder
wavelengths (the opposite of current model predictions).  A
wavelength-dependent unpolarized component could dilute the
polarization and give rise to wavelength dependence of the continuum
polarization; we cannot rule out this possibility with our data.  

\subsection{Resonant Scattering?}
\label{ssec:res_scattering}

In addition to scattering by either electrons or dust grains, resonant
scattering is a possible mechanism for the polarization observed in
the emission lines of our objects \citep{Lee1994a, Lee1994b, Lee1997}. As an example, we consider the \ion{C}{4} emission lines. Continuum or \ion{C}{4} photons emitted near the nucleus and in the dust-free gas (Figure \ref{fig:pol_model}) can resonantly scatter within the dusty BALR+BELR clouds, provided there is enough thrice-ionized carbon and the incident frequency matches the Doppler-shifted frequency of the outflowing BALR+BELR gas. The polarization fraction reaches 100\% for a single $\Delta J = 0-1$ scattering viewed at 90 degrees \citep{Hamilton1947}. 

We can calculate the expected ratio of the optical depth of resonant scattering compared to the optical depth to dust scattering. The optical depth to dust scattering, $\tau_{\rm dust}$ is given by 
\begin{equation}
\tau_{\rm dust} = N_{\rm H} \times C_{\rm ext,H}
\end{equation}
assuming a constant number density of dust grains along the line of sight. Here $N_{\rm H}$ is the column density of hydrogen nucleons and $C_{\rm ext,H}$ is the extinction cross section of the dust per hydrogen nucleon, taken to be $3.768\times10^{-22}$ cm$^{2}$/H for an SMC-like dust distribution at 1200 \AA~from \citet{Draine2003}. 

To calculate the optical depth to resonant scattering, we start with the expression for the rate of resonant excitation per atom from \citet{Lee1994a} and find 
\begin{equation}
\tau_{\rm res} = \frac{\pi e^2 f_{\rm abs}}{m_e v_{\rm wind} \nu_0} N_{\rm H} A_{\rm C} \eta_{\rm CIV}.
\end{equation}
Here $f_{\rm abs}=0.2303$ \citep{Morton1991} is the absorption strength the \ion{C}{4}$\lambda$1550\AA\ transition, $v_{\rm wind} \approx 3000$ km s$^{-1}$ is the range of velocities in the outflow, $\nu_0=c/1550$\AA\ is the rest-frame resonant frequency, $A_{\rm C}=4\times 10^{-4}$ is the carbon abundance by number \citep{Morton1991} and $\eta_{\rm CIV}$ is the fraction of carbon in the relevant ionization state. 

The biggest uncertainty in our estimate of the resonant scattering optical depth is the ionization state of \ion{C}{4}, with $\tau_{\rm res}/\tau_{\rm dust}\simeq 300\times \eta_{\rm CIV}$. In a standard dust-free BLR $\eta_{\rm CIV}\simeq 0.3$ (D. Proga, private communication), and if that were the case we would conclude that resonant scattering must be the dominant process. But in the dusty BELR+BALR region this fraction could be strongly suppressed and strongly vary as a function of distance from the inner edge of the obscuration. Therefore, we leave the question of the dominant scattering process open. As we see in the next section, the salient features of our proposed model depend most critically on the geometry of scattering and less so on the precise scattering mechanism.  

\subsection{Structure of Line Polarization}
\label{ssec:disc_linepol}

In low-redshift type 2 quasars, scattering occurs on a wide range of scales, reaching several kpc \citep{Hines1999,Zakamska2005,Zakamska2006,Schmidt2007,Obied2016}, comparable to the scales on which optical forbidden lines such as [\ion{O}{3}] are produced. The co-spatial distribution of the line-emitting gas and of the scattering medium leads to a geometric cancellation of the polarization of the emission lines and to mixing of direct (unpolarized) emission with scattered light, resulting in the observed low fractional polarization of the narrow emission-line region \citep{Stockman1981,Glenn1994,Goodrich1995b,Tran1995c}.

In our objects, we observe a lower net polarization of the UV emission lines than the continuum. By analogy to the suppression of polarization of the narrow-line region of type 2 quasars, we hypothesize that the UV emission lines arise on scales similar to those that dominate scattering (though as already discussed in Section \S\ref{ssec:line_pol}, the contribution of resonant scattering complicates this picture of geometric dilution). In the cartoon shown in Figure \ref{fig:pol_model}, the low fractional polarization of the emission lines would be achieved if the lines are produced either in the dusty BELR / BALR or in the dust-free cone at scales similar to or greater than the scales of the dusty BELR / BALR. There is also empirical evidence \citep{Czerny2011} and emission-line calculations \citep{Netzer1993} that suggest the broad-line region and inner edge of the obscuring material should exist on similar scales. Finally, in a recent paper, \citet{Baskin2018} present detailed calculations of the geometry of both the dust-free and the dusty BELR.

In the simple case of a spherical emitting region of radius $R$ and scattering region of radius $D$, \citet{Cassinelli1987} show that the polarization is reduced by a factor of $[1-(R/D)^2]^{1/2}$ relative to an emitting point source. If we assume that the continuum emitting region is a point source, and it is roughly $\sim 3\times$ more polarized than the emission-line region, this implies that the emission-line region is 94\% the size of the scattering region, confirming that the emission-line region and scattering regions extend to similar scales.

What constraints on the physical conditions do we get from requiring that continuum scattering and UV emission-line production occur on similar scales? The ionization parameter is
\begin{equation}
U=\frac{L_{\rm ion}}{4\pi r^2 \phi n_e c}.
\label{eq:ionization_parameter}
\end{equation}
Here $L_{\rm ion}$ is the luminosity of the ionizing radiation, $\phi=13.6$ eV is the threshold energy of the ionizing photons, and $n_e\simeq (1-1.2)\times n_{\rm H}$ is the electron density which depends slightly on whether helium is fully ionized. As a fiducial value of the ionization parameter in the \ion{C}{4}-emitting region we use $\log U=-0.5$ from \citet{Rodriguez2011}. As discussed in \S~\ref{ssec:line_pol}, emission lines that we see in the rest-frame UV spectra originate on scales similar to scattering -- i.e., on scales that are larger than the dust sublimation distance. Therefore, we consider the possibility that the line-emitting clouds are subjected to an ionizing continuum which is strongly suppressed by dust extinction. We plot the resulting relationship between $n_H$ and $r$ in Figure \ref{fig:dust_constraints} for various values of $L_{\rm ion}$.

It is clear from Figure \ref{fig:dust_constraints} that if emission lines are produced on spatial scales that are similar to those of scattering, then either a strong suppression of the ionizing continuum or strong clumping of the emission-line gas, or both, are required to produce the observed ionization parameter. We will model the UV emission lines of ERQs in a forthcoming photo-ionization analysis of the emission-line ratios (Hamann et al., in prep.). In the meanwhile, our qualitative model is that the emission lines are produced in high density clouds either in the joint BELR / BALR or in the dust-free zone. These clouds are ionization-bounded and therefore are by definition optically thick to ionizing radiation, so only the compact surfaces of these clouds will be able to scatter continuum emission in the dusty zone. Thus, scattering is dominated by a lower-density, volume-filling component of the BELR / BALR gas. This allows us to reconcile the densities implied by the scattering efficiency and by the ionization parameter (see Figure \ref{fig:dust_constraints}).

The biggest success of our model in Figure \ref{fig:pol_model} is that it naturally explains the change in the polarization fraction and position angle as a function of velocity in the FUV emission lines. The blueshifted line emission we see in our objects is from the near side of the equatorial outflow which gives it the observed velocity shift (see the right panel of Figure \ref{fig:pol_model}). In contrast, the observed redshifted emission is dominated not by the back of the outflow (this emission is suppressed because dust back scattering is inefficient) but instead from the sides of the equatorial outflow which are moving away from the observer and thus produce redshifted scattered emission. This results in the swing in polarization position angle as a function of velocity in the emission lines, as the two scattering regions are roughly perpendicular to each other as seen in the plane of the sky. As a result, the polarization position angle transitions as a function of velocity from being dominated by the near-side of the equatorial outflow in the blue to the sides of the outflow in the red. This also explains the overall redshift in emission-line features seen in polarized light as the net redshift of the scattered line is due to the redshift of the scatterer relative to the emitter. Furthermore, the sides of the outflow dominate the net scattering of the continuum, and we expect the red-wing polarization position angle and the continuum position angle to match.

This explanation for the polarization position angle swing is independent of the scattering mechanism dominating within the lines -- whether it is dust, resonant scattering or (unlikely) electrons. In all three mechanisms, the dominant polarization position angle is orthogonal to the dominant scattering direction as projected on the plane of the sky. Thus, the key features of our model -- the 90 degree rotation of the polarization position angle and the net line redshift -- reflect the geometry and the kinematics of the equatorial outflow.

If the scattering is dominated by dust, then our model predicts the polarization fraction of the blue wing to be low due to the low polarization efficiency of forward scattering. This is indeed clearly seen in the data. In Figure \ref{fig:velocity}, the dominant polarization signal is represented by the $u$ values. The absolute value of $u$ in all three lines reaches 30\% at $v\simeq +1500$ km sec$^{-1}$ and only 5\% on the blue end at $v<0$. We leave a full examination of dust+resonant scattering models for future work. 

\citet{Smith2004} suggested a similar model of the torus with both polar and equatorial scattering regions to explain the range of properties observed in low redshift Seyfert 1s and 2s. The polar scatterer extends beyond the torus and may be outflowing while the equatorial scatterer lies within the torus, and is described by \citet{Young2000} as a rotating disk wind.  Other differences between our model and the \citet{Young2000} model include the specific emission lines: they consider Balmer lines, which in our cartoon are produced in a more extended region of the disk than the FUV lines which we focus on. Furthermore, in Type 1 objects there is dilution of the scattered light by the direct unpolarized continuum, which may explain the much lower values of polarization and position angle swings achieved in Seyfert 1s. But the main difference is that \citet{Young2000} explains the rotation in the polarization position angle using a rotating disk wind, where rotation might be difficult to maintain dynamically. In contrast, in our model the polarization position angle swing is achieved naturally by the geometric projection effects, with the blueshifted scattered emission and the redshifted scattered emission having different orientations in projection on the plane of the sky. 

\section{Conclusions}
\label{sec:conclusions}

We have obtained rest-frame ultraviolet spectropolarimetry for five $z \sim 2.5$ obscured and highly reddened quasars using LRIS in polarimetry mode on Keck.  Our Type 2 targets are classically selected, narrow-line objects \citep{Alexandroff2013}. Extremely red quasars are color-selected and show signs of extreme outflow activity in their rest-frame optical spectra \citep{Ross2015,Zakamska2016,Hamann2017}.   Although our initial expectations were that narrow line-selected Type 2 quasars should have a more axisymmetric scattering region and that they should therefore be more highly polarized than the extremely red quasars, we find (in this admittedly small sample) no systematic trends between the optical and UV ``type" and the levels of continuum polarization.  

We do see several interesting trends in our polarization data:
\begin{enumerate}
\item We find high levels of polarization in the continuum, higher than $10\%$ in three of our sources.
\item We find lower levels of polarization in emission lines with a ratio of emission-line polarization/continuum polarization between 0.3 and 0.8.  
\item Intriguingly, we see a rotation of almost exactly ninety degrees in the polarization position angle across all strong emission lines, Ly$\alpha$, \ion{C}{4}, and \ion{N}{5}; 
\item The polarization position angle of the continuum matches the polarization position angle in the redshifted wings of the emission lines.
\end{enumerate}

To explain our data, we present a model in which an equatorial dusty
disk wind scatters both continuum emission from the disk and line
emission from a polar outflow. This model is physically motivated and
appears to match the main features of our observations.  The high
polarization fraction of the continuum can be explained by dust scattering, although
existing dust models may need further refinement.  The fact that the
polarization is lower in the emission lines than the continuum
suggests that the emission-line region and dusty scattering region
extend over similar scales.  Finally, the model explains the observed
$90^\circ$ rotation of the polarization position angle through
emission lines by invoking scattering off of kinematically distinct
regions of the emission-line region. 

Spectropolarimetry is an important tool to understand the scattering
geometry of our high redshift obscured and reddened quasars and can be
an important piece of information to evaluate different proposed
models of quasar geometry on small scales near the black hole. While
polarimetry measurements remain difficult, searching for a small
signal in already faint objects, it appears that obscured and red
quasars at redshifts where host galaxy contamination is small often
show polarization fractions in excess of 10\%, which makes the
prospect bright for future studies of this population. Future spectropolarimetric studies will allow us a window into the potential launching mechanism for galaxy-scale quasar winds which are a necessary ingredient for quasar feedback. 

\acknowledgements

The data presented herein were obtained at the W.M. Keck Observatory, which is operated as a scientific partnership among the California Institute of Technology, the University of California and the National Aeronautics and Space Administration. The Observatory was made possible by the generous financial support of the W.M. Keck Foundation.  The authors wish to recognize and acknowledge the very significant cultural role and reverence that the summit of Maunakea has always had within the indigenous Hawaiian community.  We are most fortunate to have the opportunity to conduct observations from this mountain.

We would like to thank the referee for many helpful comments and suggestions the inclusion of which helped to improve our paper. R.A. would like to acknowledge the assistance of C. Steidel, A. Strom, D. Perley, and H. Tran during observations at Keck I and the assistance of M. Kassis on the evening of LRISp observations. R.A. would also like to thank D. Neufeld for useful conversations and S. Veilleux for allowing us to adapt figure 8 of \citet{Veilleux2016}. N.L.Z. would like to acknowledge the remote observing support of Yale University.

R.A. was supported in part by NASA JPL grant 1520456.  Support for this work was provided in part by the National Aeronautics and Space Administration through Chandra Award Number GO6-17100X issued by the Chandra X-ray Observatory Center, which is operated by the Smithsonian Astrophysical Observatory for and on behalf of the National Aeronautics Space Administration under contract NAS8-03060. N.L.Z. acknowledges support by the Catalyst award of the Johns Hopkins University. Research by A.J.B. is supported in part by NSF grant AST-1412693.


\begin{thebibliography}{}
\expandafter\ifx\csname natexlab\endcsname\relax\def\natexlab#1{#1}\fi

\bibitem[{{Abramowicz} {et~al.}(1988){Abramowicz}, {Czerny}, {Lasota}, \&
  {Szuszkiewicz}}]{Abramowicz1988}
{Abramowicz}, M.~A., {Czerny}, B., {Lasota}, J.~P., \& {Szuszkiewicz}, E. 1988,
  \apj, 332, 646

\bibitem[{{Ahn} {et~al.}(2012){Ahn}, {Alexandroff}, {Allende Prieto},
  {Anderson}, {Anderton}, {Andrews}, {Aubourg}, {Bailey}, {Balbinot}, {Barnes},
  \& et~al.}]{Ahn2012}
{Ahn}, C.~P., {Alexandroff}, R., {Allende Prieto}, C., {et~al.} 2012, \apjs,
  203, 21

\bibitem[{{Alexandroff} {et~al.}(2013){Alexandroff}, {Strauss}, {Greene},
  {Zakamska}, {Ross}, {Brandt}, {Liu}, {Smith}, {Ge}, {Hamann}, {Myers},
  {Petitjean}, {Schneider}, {Yesuf}, \& {York}}]{Alexandroff2013}
{Alexandroff}, R., {Strauss}, M.~A., {Greene}, J.~E., {et~al.} 2013, \mnras,
  435, 3306

\bibitem[{{Antonucci}(1993)}]{Antonucci1993}
{Antonucci}, R. 1993, \araa, 31, 473

\bibitem[{{Antonucci} \& {Miller}(1985)}]{Antonucci1985}
{Antonucci}, R.~R.~J., \& {Miller}, J.~S. 1985, \apj, 297, 621

\bibitem[{{Assef} {et~al.}(2016){Assef}, {Walton}, {Brightman}, {Stern},
  {Alexander}, {Bauer}, {Blain}, {Diaz-Santos}, {Eisenhardt}, {Finkelstein},
  {Hickox}, {Tsai}, \& {Wu}}]{Assef2016}
{Assef}, R.~J., {Walton}, D.~J., {Brightman}, M., {et~al.} 2016, \apj, 819, 111

\bibitem[{{Banerji} {et~al.}(2015){Banerji}, {Alaghband-Zadeh}, {Hewett}, \&
  {McMahon}}]{Banerji2015}
{Banerji}, M., {Alaghband-Zadeh}, S., {Hewett}, P.~C., \& {McMahon}, R.~G.
  2015, \mnras, 447, 3368

\bibitem[{{Barth} {et~al.}(1999){Barth}, {Filippenko}, \& {Moran}}]{Barth1999}
{Barth}, A.~J., {Filippenko}, A.~V., \& {Moran}, E.~C. 1999, \apj, 525, 673

\bibitem[{{Barvainis}(1987)}]{Barvainis1987}
{Barvainis}, R. 1987, \apj, 320, 537

\bibitem[{{Baskin} \& {Laor}(2018)}]{Baskin2018}
{Baskin}, A., \& {Laor}, A. 2018, \mnras, 474, 1970

\bibitem[{{Berriman} {et~al.}(1990){Berriman}, {Schmidt}, {West}, \&
  {Stockman}}]{Berriman1990}
{Berriman}, G., {Schmidt}, G.~D., {West}, S.~C., \& {Stockman}, H.~S. 1990,
  \apjs, 74, 869

\bibitem[{{Borguet} {et~al.}(2008){Borguet}, {Hutsem{\'e}kers}, {Letawe},
  {Letawe}, \& {Magain}}]{Borguet2008}
{Borguet}, B., {Hutsem{\'e}kers}, D., {Letawe}, G., {Letawe}, Y., \& {Magain},
  P. 2008, \aap, 478, 321

\bibitem[{{Cassinelli} {et~al.}(1987){Cassinelli}, {Nordsieck}, \&
  {Murison}}]{Cassinelli1987}
{Cassinelli}, J.~P., {Nordsieck}, K.~H., \& {Murison}, M.~A. 1987, \apj, 317,
  290

\bibitem[{{Chan} \& {Krolik}(2016)}]{Chan2016}
{Chan}, C.-H., \& {Krolik}, J.~H. 2016, \apj, 825, 67

\bibitem[{{Clemens} \& {Tapia}(1990)}]{Clemens1990}
{Clemens}, D.~P., \& {Tapia}, S. 1990, \pasp, 102, 179

\bibitem[{{Czerny} \& {Hryniewicz}(2011)}]{Czerny2011}
{Czerny}, B., \& {Hryniewicz}, K. 2011, \aap, 525, L8

\bibitem[{{Dawson} {et~al.}(2013){Dawson}, {Schlegel}, {Ahn}, {Anderson},
  {Aubourg}, {Bailey}, {Barkhouser}, {Bautista}, {Beifiori}, {Berlind}, \&
  et~al.}]{Dawson2013}
{Dawson}, K.~S., {Schlegel}, D.~J., {Ahn}, C.~P., {et~al.} 2013, \aj, 145, 10

\bibitem[{{Dey} {et~al.}(1996){Dey}, {Cimatti}, {van Breugel}, {Antonucci}, \&
  {Spinrad}}]{Dey1996}
{Dey}, A., {Cimatti}, A., {van Breugel}, W., {Antonucci}, R., \& {Spinrad}, H.
  1996, \apj, 465, 157

\bibitem[{{Donley} {et~al.}(2012){Donley}, {Koekemoer}, {Brusa}, {Capak},
  {Cardamone}, {Civano}, {Ilbert}, {Impey}, {Kartaltepe}, {Miyaji}, {Salvato},
  {Sanders}, {Trump}, \& {Zamorani}}]{Donley2012}
{Donley}, J.~L., {Koekemoer}, A.~M., {Brusa}, M., {et~al.} 2012, \apj, 748, 142

\bibitem[{{Draine}(2003)}]{Draine2003}
{Draine}, B.~T. 2003, \apj, 598, 1017

\bibitem[{{Eisenhardt} {et~al.}(2012){Eisenhardt}, {Wu}, {Tsai}, {Assef},
  {Benford}, {Blain}, {Bridge}, {Condon}, {Cushing}, {Cutri}, {Evans},
  {Gelino}, {Griffith}, {Grillmair}, {Jarrett}, {Lonsdale}, {Masci}, {Mason},
  {Petty}, {Sayers}, {Stanford}, {Stern}, {Wright}, \& {Yan}}]{Eisenhardt2012}
{Eisenhardt}, P.~R.~M., {Wu}, J., {Tsai}, C.-W., {et~al.} 2012, \apj, 755, 173

\bibitem[{{Elitzur} \& {Shlosman}(2006)}]{Elitzur2006}
{Elitzur}, M., \& {Shlosman}, I. 2006, \apjl, 648, L101

\bibitem[{{Elvis}(2017)}]{Elvis2017}
{Elvis}, M. 2017, ArXiv e-prints, arXiv:1703.02956

\bibitem[{{Glenn} {et~al.}(1994){Glenn}, {Schmidt}, \& {Foltz}}]{Glenn1994}
{Glenn}, J., {Schmidt}, G.~D., \& {Foltz}, C.~B. 1994, \apjl, 434, L47

\bibitem[{{Glikman} {et~al.}(2007){Glikman}, {Helfand}, {White}, {Becker},
  {Gregg}, \& {Lacy}}]{Glikman2007}
{Glikman}, E., {Helfand}, D.~J., {White}, R.~L., {et~al.} 2007, \apj, 667, 673

\bibitem[{{Glikman} {et~al.}(2015){Glikman}, {Simmons}, {Mailly}, {Schawinski},
  {Urry}, \& {Lacy}}]{Glikman2015}
{Glikman}, E., {Simmons}, B., {Mailly}, M., {et~al.} 2015, \apj, 806, 218

\bibitem[{{Glikman} {et~al.}(2012){Glikman}, {Urrutia}, {Lacy}, {Djorgovski},
  {Mahabal}, {Myers}, {Ross}, {Petitjean}, {Ge}, {Schneider}, \&
  {York}}]{Glikman2012}
{Glikman}, E., {Urrutia}, T., {Lacy}, M., {et~al.} 2012, \apj, 757, 51

\bibitem[{{Glikman} {et~al.}(2013){Glikman}, {Urrutia}, {Lacy}, {Djorgovski},
  {Urry}, {Croom}, {Schneider}, {Mahabal}, {Graham}, \& {Ge}}]{Glikman2013}
---. 2013, \apj, 778, 127

\bibitem[{{Goodrich} {et~al.}(1995){Goodrich}, {Cohen}, \&
  {Putney}}]{Goodrich1995a}
{Goodrich}, R.~W., {Cohen}, M.~H., \& {Putney}, A. 1995, \pasp, 107, 179

\bibitem[{{Goodrich} \& {Miller}(1995)}]{Goodrich1995b}
{Goodrich}, R.~W., \& {Miller}, J.~S. 1995, \apjl, 448, L73

\bibitem[{{Goulding} {et~al.}(2018){Goulding}, {Zakamska}, {Alexandroff},
  {Assef}, {Banerji}, {Hamann}, {Wylezalek}, {Brandt}, {Greene}, {Lansbury},
  {P{\^a}ris}, {Richards}, {Stern}, \& {Strauss}}]{goul18}
{Goulding}, A.~D., {Zakamska}, N.~L., {Alexandroff}, R.~M., {et~al.} 2018,
  \apj, 856, 4

\bibitem[{{Greene} {et~al.}(2014){Greene}, {Alexandroff}, {Strauss},
  {Zakamska}, {Lang}, {Liu}, {Pattarakijwanich}, {Hamann}, {Ross}, {Myers},
  {Brandt}, {York}, \& {Schneider}}]{Greene2014}
{Greene}, J.~E., {Alexandroff}, R., {Strauss}, M.~A., {et~al.} 2014, \apj, 788,
  91

\bibitem[{{Hamann} {et~al.}(2011){Hamann}, {Kanekar}, {Prochaska}, {Murphy},
  {Ellison}, {Malec}, {Milutinovic}, \& {Ubachs}}]{hama11}
{Hamann}, F., {Kanekar}, N., {Prochaska}, J.~X., {et~al.} 2011, \mnras, 410,
  1957

\bibitem[{{Hamann} {et~al.}(2017){Hamann}, {Zakamska}, {Ross}, {Paris},
  {Alexandroff}, {Villforth}, {Richards}, {Herbst}, {Brandt}, {Cook}, {Denney},
  {Greene}, {Schneider}, \& {Strauss}}]{Hamann2017}
{Hamann}, F., {Zakamska}, N.~L., {Ross}, N., {et~al.} 2017, \mnras, 464, 3431

\bibitem[{{Hamilton}(1947)}]{Hamilton1947}
{Hamilton}, D.~R. 1947, \apj, 106, 457

\bibitem[{{Hao} {et~al.}(2005){Hao}, {Strauss}, {Tremonti}, {Schlegel},
  {Heckman}, {Kauffmann}, {Blanton}, {Fan}, {Gunn}, {Hall}, {Ivezi{\'c}},
  {Knapp}, {Krolik}, {Lupton}, {Richards}, {Schneider}, {Strateva}, {Zakamska},
  {Brinkmann}, {Brunner}, \& {Szokoly}}]{Hao2005}
{Hao}, L., {Strauss}, M.~A., {Tremonti}, C.~A., {et~al.} 2005, \aj, 129, 1783

\bibitem[{{Hines} {et~al.}(2001){Hines}, {Schmidt}, {Gordon}, {Smith}, {Wills},
  {Allen}, \& {Sitko}}]{Hines2001}
{Hines}, D.~C., {Schmidt}, G.~D., {Gordon}, K.~D., {et~al.} 2001, \apj, 563,
  512

\bibitem[{{Hines} {et~al.}(1995){Hines}, {Schmidt}, {Smith}, {Cutri}, \&
  {Low}}]{Hines1995}
{Hines}, D.~C., {Schmidt}, G.~D., {Smith}, P.~S., {Cutri}, R.~M., \& {Low},
  F.~J. 1995, \apjl, 450, L1

\bibitem[{{Hines} {et~al.}(1999){Hines}, {Schmidt}, {Wills}, {Smith}, \&
  {Sowinski}}]{Hines1999}
{Hines}, D.~C., {Schmidt}, G.~D., {Wills}, B.~J., {Smith}, P.~S., \&
  {Sowinski}, L.~G. 1999, \apj, 512, 145

\bibitem[{{Hoeflich} {et~al.}(1996){Hoeflich}, {Wheeler}, {Hines}, \&
  {Trammell}}]{Hoeflich1996}
{Hoeflich}, P., {Wheeler}, J.~C., {Hines}, D.~C., \& {Trammell}, S.~R. 1996,
  \apj, 459, 307

\bibitem[{{Hopkins} {et~al.}(2006){Hopkins}, {Hernquist}, {Cox}, {Di Matteo},
  {Robertson}, \& {Springel}}]{Hopkins2006}
{Hopkins}, P.~F., {Hernquist}, L., {Cox}, T.~J., {et~al.} 2006, \apjs, 163, 1

\bibitem[{{Horne}(1986)}]{Horne1986}
{Horne}, K. 1986, \pasp, 98, 609

\bibitem[{{Kasen} {et~al.}(2003){Kasen}, {Nugent}, {Wang}, {Howell}, {Wheeler},
  {H{\"o}flich}, {Baade}, {Baron}, \& {Hauschildt}}]{Kasen2003}
{Kasen}, D., {Nugent}, P., {Wang}, L., {et~al.} 2003, \apj, 593, 788

\bibitem[{{Kauffmann} {et~al.}(2003){Kauffmann}, {Heckman}, {Tremonti},
  {Brinchmann}, {Charlot}, {White}, {Ridgway}, {Brinkmann}, {Fukugita}, {Hall},
  {Ivezi{\'c}}, {Richards}, \& {Schneider}}]{Kauffmann2003}
{Kauffmann}, G., {Heckman}, T.~M., {Tremonti}, C., {et~al.} 2003, \mnras, 346,
  1055

\bibitem[{{Khachikian} \& {Weedman}(1974)}]{khac74}
{Khachikian}, E.~Y., \& {Weedman}, D.~W. 1974, \apj, 192, 581

\bibitem[{{Kishimoto} {et~al.}(2001){Kishimoto}, {Antonucci}, {Cimatti},
  {Hurt}, {Dey}, {van Breugel}, \& {Spinrad}}]{Kishimoto2001}
{Kishimoto}, M., {Antonucci}, R., {Cimatti}, A., {et~al.} 2001, \apj, 547, 667

\bibitem[{{Lee}(1994)}]{Lee1994b}
{Lee}, H.~W. 1994, \mnras, 268, 49

\bibitem[{{Lee} \& {Blandford}(1997)}]{Lee1997}
{Lee}, H.-W., \& {Blandford}, R.~D. 1997, \mnras, 288, 19

\bibitem[{{Lee} {et~al.}(1994){Lee}, {Blandford}, \& {Western}}]{Lee1994a}
{Lee}, H.-W., {Blandford}, R.~D., \& {Western}, L. 1994, \mnras, 267, 303

\bibitem[{{Mason} {et~al.}(2015){Mason}, {Rodr{\'{\i}}guez-Ardila}, {Martins},
  {Riffel}, {Gonz{\'a}lez Mart{\'{\i}}n}, {Ramos Almeida}, {Ruschel Dutra},
  {Ho}, {Thanjavur}, {Flohic}, {Alonso-Herrero}, {Lira}, {McDermid}, {Riffel},
  {Schiavon}, {Winge}, {Hoenig}, \& {Perlman}}]{Mason2015}
{Mason}, R.~E., {Rodr{\'{\i}}guez-Ardila}, A., {Martins}, L., {et~al.} 2015,
  \apjs, 217, 13

\bibitem[{{McKinney} {et~al.}(2015){McKinney}, {Dai}, \&
  {Avara}}]{McKinney2015}
{McKinney}, J.~C., {Dai}, L., \& {Avara}, M.~J. 2015, \mnras, 454, L6

\bibitem[{{Miller} \& {Goodrich}(1990)}]{Miller1990}
{Miller}, J.~S., \& {Goodrich}, R.~W. 1990, \apj, 355, 456

\bibitem[{{Miller} {et~al.}(1991){Miller}, {Goodrich}, \&
  {Mathews}}]{Miller1991}
{Miller}, J.~S., {Goodrich}, R.~W., \& {Mathews}, W.~G. 1991, \apj, 378, 47

\bibitem[{{Miller} {et~al.}(1988){Miller}, {Robinson}, \&
  {Goodrich}}]{Miller1988}
{Miller}, J.~S., {Robinson}, L.~B., \& {Goodrich}, R.~W. 1988, in
  Instrumentation for Ground-Based Optical Astronomy, ed. L.~B. {Robinson}, 157

\bibitem[{{Morton}(1991)}]{Morton1991}
{Morton}, D.~C. 1991, The Astrophysical Journal Supplement Series, 77, 119

\bibitem[{{Netzer} \& {Laor}(1993)}]{Netzer1993}
{Netzer}, H., \& {Laor}, A. 1993, \apjl, 404, L51

\bibitem[{{Norman} {et~al.}(2002){Norman}, {Hasinger}, {Giacconi}, {Gilli},
  {Kewley}, {Nonino}, {Rosati}, {Szokoly}, {Tozzi}, {Wang}, {Zheng}, {Zirm},
  {Bergeron}, {Gilmozzi}, {Grogin}, {Koekemoer}, \& {Schreier}}]{Norman2002}
{Norman}, C., {Hasinger}, G., {Giacconi}, R., {et~al.} 2002, \apj, 571, 218

\bibitem[{{Obied} {et~al.}(2016){Obied}, {Zakamska}, {Wylezalek}, \&
  {Liu}}]{Obied2016}
{Obied}, G., {Zakamska}, N.~L., {Wylezalek}, D., \& {Liu}, G. 2016, \mnras,
  456, 2861

\bibitem[{{Oke} {et~al.}(1995){Oke}, {Cohen}, {Carr}, {Cromer}, {Dingizian},
  {Harris}, {Labrecque}, {Lucinio}, {Schaal}, {Epps}, \& {Miller}}]{Oke1995}
{Oke}, J.~B., {Cohen}, J.~G., {Carr}, M., {et~al.} 1995, \pasp, 107, 375

\bibitem[{{Reyes} {et~al.}(2008){Reyes}, {Zakamska}, {Strauss}, {Green},
  {Krolik}, {Shen}, {Richards}, {Anderson}, \& {Schneider}}]{Reyes2008}
{Reyes}, R., {Zakamska}, N.~L., {Strauss}, M.~A., {et~al.} 2008, \aj, 136, 2373

\bibitem[{{Reynolds} {et~al.}(2013){Reynolds}, {Punsly}, {O'Dea}, \&
  {Hurley-Walker}}]{Reynolds2013}
{Reynolds}, C., {Punsly}, B., {O'Dea}, C.~P., \& {Hurley-Walker}, N. 2013,
  \apjl, 776, L21

\bibitem[{{Richards} {et~al.}(2006){Richards}, {Lacy}, {Storrie-Lombardi},
  {Hall}, {Gallagher}, {Hines}, {Fan}, {Papovich}, {Vanden Berk}, {Trammell},
  {Schneider}, {Vestergaard}, {York}, {Jester}, {Anderson}, {Budav{\'a}ri}, \&
  {Szalay}}]{Richards2006}
{Richards}, G.~T., {Lacy}, M., {Storrie-Lombardi}, L.~J., {et~al.} 2006, \apjs,
  166, 470

\bibitem[{{Rockosi} {et~al.}(2010){Rockosi}, {Stover}, {Kibrick}, {Lockwood},
  {Peck}, {Cowley}, {Bolte}, {Adkins}, {Alcott}, {Allen}, {Brown}, {Cabak},
  {Deich}, {Hilyard}, {Kassis}, {Lanclos}, {Lewis}, {Pfister}, {Phillips},
  {Robinson}, {Saylor}, {Thompson}, {Ward}, {Wei}, \& {Wright}}]{Rockosi2010}
{Rockosi}, C., {Stover}, R., {Kibrick}, R., {et~al.} 2010, in \procspie, Vol.
  7735, Ground-based and Airborne Instrumentation for Astronomy III, 77350R

\bibitem[{{Rodr{\'{\i}}guez Hidalgo} {et~al.}(2011){Rodr{\'{\i}}guez Hidalgo},
  {Hamann}, \& {Hall}}]{Rodriguez2011}
{Rodr{\'{\i}}guez Hidalgo}, P., {Hamann}, F., \& {Hall}, P. 2011, \mnras, 411,
  247

\bibitem[{{Ross} {et~al.}(2015){Ross}, {Hamann}, {Zakamska}, {Richards},
  {Villforth}, {Strauss}, {Greene}, {Alexandroff}, {Brandt}, {Liu}, {Myers},
  {P{\^a}ris}, \& {Schneider}}]{Ross2015}
{Ross}, N.~P., {Hamann}, F., {Zakamska}, N.~L., {et~al.} 2015, \mnras, 453,
  3932

\bibitem[{{Sanders} {et~al.}(1988){Sanders}, {Soifer}, {Elias}, {Madore},
  {Matthews}, {Neugebauer}, \& {Scoville}}]{Sanders1988}
{Sanders}, D.~B., {Soifer}, B.~T., {Elias}, J.~H., {et~al.} 1988, \apj, 325, 74

\bibitem[{{S{\c a}dowski} {et~al.}(2015){S{\c a}dowski}, {Narayan},
  {Tchekhovskoy}, {Abarca}, {Zhu}, \& {McKinney}}]{Scadowski2015}
{S{\c a}dowski}, A., {Narayan}, R., {Tchekhovskoy}, A., {et~al.} 2015, \mnras,
  447, 49

\bibitem[{{Schlafly} \& {Finkbeiner}(2011)}]{Schlafly2011}
{Schlafly}, E.~F., \& {Finkbeiner}, D.~P. 2011, \apj, 737, 103

\bibitem[{{Schlegel} {et~al.}(1998){Schlegel}, {Finkbeiner}, \&
  {Davis}}]{Schlegel1998}
{Schlegel}, D.~J., {Finkbeiner}, D.~P., \& {Davis}, M. 1998, \apj, 500, 525

\bibitem[{{Schmidt} {et~al.}(1992{\natexlab{a}}){Schmidt}, {Elston}, \&
  {Lupie}}]{Schmidt1992a}
{Schmidt}, G.~D., {Elston}, R., \& {Lupie}, O.~L. 1992{\natexlab{a}}, \aj, 104,
  1563

\bibitem[{{Schmidt} {et~al.}(2007){Schmidt}, {Smith}, {Hines}, {Tremonti}, \&
  {Low}}]{Schmidt2007}
{Schmidt}, G.~D., {Smith}, P.~S., {Hines}, D.~C., {Tremonti}, C.~A., \& {Low},
  F.~J. 2007, \apj, 666, 784

\bibitem[{{Schmidt} {et~al.}(1992{\natexlab{b}}){Schmidt}, {Stockman}, \&
  {Smith}}]{Schmidt1992b}
{Schmidt}, G.~D., {Stockman}, H.~S., \& {Smith}, P.~S. 1992{\natexlab{b}},
  \apjl, 398, L57

\bibitem[{{Serkowski} {et~al.}(1975){Serkowski}, {Mathewson}, \&
  {Ford}}]{Serkowski1975}
{Serkowski}, K., {Mathewson}, D.~S., \& {Ford}, V.~L. 1975, \apj, 196, 261

\bibitem[{{Smith} {et~al.}(2004){Smith}, {Robinson}, {Alexander}, {Young},
  {Axon}, \& {Corbett}}]{Smith2004}
{Smith}, J.~E., {Robinson}, A., {Alexander}, D.~M., {et~al.} 2004, \mnras, 350,
  140

\bibitem[{{Smith} {et~al.}(2005){Smith}, {Robinson}, {Young}, {Axon}, \&
  {Corbett}}]{Smith2005}
{Smith}, J.~E., {Robinson}, A., {Young}, S., {Axon}, D.~J., \& {Corbett}, E.~A.
  2005, \mnras, 359, 846

\bibitem[{{Smith} {et~al.}(1995){Smith}, {Schmidt}, {Allen}, \&
  {Angel}}]{Smith1995}
{Smith}, P.~S., {Schmidt}, G.~D., {Allen}, R.~G., \& {Angel}, J.~R.~P. 1995,
  \apj, 444, 146

\bibitem[{{Smith} {et~al.}(2000){Smith}, {Schmidt}, {Hines}, {Cutri}, \&
  {Nelson}}]{Smith2000}
{Smith}, P.~S., {Schmidt}, G.~D., {Hines}, D.~C., {Cutri}, R.~M., \& {Nelson},
  B.~O. 2000, \apjl, 545, L19

\bibitem[{{Smith} {et~al.}(2002){Smith}, {Schmidt}, {Hines}, {Cutri}, \&
  {Nelson}}]{Smith2002}
---. 2002, \apj, 569, 23

\bibitem[{{Smith} {et~al.}(2003){Smith}, {Schmidt}, {Hines}, \&
  {Foltz}}]{Smith2003}
{Smith}, P.~S., {Schmidt}, G.~D., {Hines}, D.~C., \& {Foltz}, C.~B. 2003, \apj,
  593, 676

\bibitem[{{Stern} {et~al.}(2002){Stern}, {Moran}, {Coil}, {Connolly}, {Davis},
  {Dawson}, {Dey}, {Eisenhardt}, {Elston}, {Graham}, {Harrison}, {Helfand},
  {Holden}, {Mao}, {Rosati}, {Spinrad}, {Stanford}, {Tozzi}, \&
  {Wu}}]{Stern2002}
{Stern}, D., {Moran}, E.~C., {Coil}, A.~L., {et~al.} 2002, \apj, 568, 71

\bibitem[{{Stern} {et~al.}(2005){Stern}, {Eisenhardt}, {Gorjian}, {Kochanek},
  {Caldwell}, {Eisenstein}, {Brodwin}, {Brown}, {Cool}, {Dey}, {Green},
  {Jannuzi}, {Murray}, {Pahre}, \& {Willner}}]{Stern2005}
{Stern}, D., {Eisenhardt}, P., {Gorjian}, V., {et~al.} 2005, \apj, 631, 163

\bibitem[{{Stockman} {et~al.}(1981){Stockman}, {Hier}, \&
  {Angel}}]{Stockman1981}
{Stockman}, H.~S., {Hier}, R.~G., \& {Angel}, J.~R.~P. 1981, \apj, 243, 404

\bibitem[{{Tadhunter}(2005)}]{Tadhunter2005}
{Tadhunter}, C. 2005, in Astronomical Society of the Pacific Conference Series,
  Vol. 343, Astronomical Polarimetry: Current Status and Future Directions, ed.
  A.~{Adamson}, C.~{Aspin}, C.~{Davis}, \& T.~{Fujiyoshi}, 457

\bibitem[{{Tran}(1995)}]{Tran1995c}
{Tran}, H.~D. 1995, \apj, 440, 597

\bibitem[{{Tran} {et~al.}(1995){Tran}, {Cohen}, \& {Goodrich}}]{Tran1995a}
{Tran}, H.~D., {Cohen}, M.~H., \& {Goodrich}, R.~W. 1995, \aj, 110, 2597

\bibitem[{{Treister} {et~al.}(2009){Treister}, {Cardamone}, {Schawinski},
  {Urry}, {Gawiser}, {Virani}, {Lira}, {Kartaltepe}, {Damen}, {Taylor}, {Le
  Floc'h}, {Justham}, \& {Koekemoer}}]{Treister2009}
{Treister}, E., {Cardamone}, C.~N., {Schawinski}, K., {et~al.} 2009, \apj, 706,
  535

\bibitem[{{Tsai} {et~al.}(2015){Tsai}, {Eisenhardt}, {Wu}, {Stern}, {Assef},
  {Blain}, {Bridge}, {Benford}, {Cutri}, {Griffith}, {Jarrett}, {Lonsdale},
  {Masci}, {Moustakas}, {Petty}, {Sayers}, {Stanford}, {Wright}, {Yan},
  {Leisawitz}, {Liu}, {Mainzer}, {McLean}, {Padgett}, {Skrutskie}, {Gelino},
  {Beichman}, \& {Juneau}}]{Tsai2015}
{Tsai}, C.-W., {Eisenhardt}, P.~R.~M., {Wu}, J., {et~al.} 2015, \apj, 805, 90

\bibitem[{{Urrutia} {et~al.}(2009){Urrutia}, {Becker}, {White}, {Glikman},
  {Lacy}, {Hodge}, \& {Gregg}}]{Urrutia2009}
{Urrutia}, T., {Becker}, R.~H., {White}, R.~L., {et~al.} 2009, \apj, 698, 1095

\bibitem[{{Urry} \& {Padovani}(1995)}]{Urry1995}
{Urry}, C.~M., \& {Padovani}, P. 1995, \pasp, 107, 803

\bibitem[{{van Dokkum}(2001)}]{VanDokkum2001}
{van Dokkum}, P.~G. 2001, \pasp, 113, 1420

\bibitem[{{Veilleux} {et~al.}(2016){Veilleux}, {Melendez}, {Tripp}, {Hamann},
  \& {Rupke}}]{Veilleux2016}
{Veilleux}, S., {Melendez}, M., {Tripp}, T.~M., {Hamann}, F., \& {Rupke},
  D.~S.~N. 2016, ArXiv e-prints, arXiv:1605.00665

\bibitem[{{Vernet} {et~al.}(2001){Vernet}, {Fosbury}, {Villar-Mart{\'{\i}}n},
  {Cohen}, {Cimatti}, {di Serego Alighieri}, \& {Goodrich}}]{Vernet2001}
{Vernet}, J., {Fosbury}, R.~A.~E., {Villar-Mart{\'{\i}}n}, M., {et~al.} 2001,
  \aap, 366, 7

\bibitem[{{Wang} {et~al.}(2001){Wang}, {Howell}, {H{\"o}flich}, \&
  {Wheeler}}]{Wang2001}
{Wang}, L., {Howell}, D.~A., {H{\"o}flich}, P., \& {Wheeler}, J.~C. 2001, \apj,
  550, 1030

\bibitem[{{Wang} \& {Wheeler}(2008)}]{Wang2008}
{Wang}, L., \& {Wheeler}, J.~C. 2008, \araa, 46, 433

\bibitem[{{Weingartner} \& {Draine}(2001)}]{Weingartner2001}
{Weingartner}, J.~C., \& {Draine}, B.~T. 2001, \apj, 548, 296

\bibitem[{{Wills} {et~al.}(1992){Wills}, {Wills}, {Evans}, {Natta}, {Thompson},
  {Breger}, \& {Sitko}}]{Wills1992}
{Wills}, B.~J., {Wills}, D., {Evans}, II, N.~J., {et~al.} 1992, \apj, 400, 96

\bibitem[{{Wright} {et~al.}(2010){Wright}, {Eisenhardt}, {Mainzer}, {Ressler},
  {Cutri}, {Jarrett}, {Kirkpatrick}, {Padgett}, {McMillan}, {Skrutskie}, \&
  et~al.}]{White2010}
{Wright}, E.~L., {Eisenhardt}, P.~R.~M., {Mainzer}, A.~K., {et~al.} 2010, \aj,
  140, 1868

\bibitem[{{Wu} {et~al.}(2012){Wu}, {Tsai}, {Sayers}, {Benford}, {Bridge},
  {Blain}, {Eisenhardt}, {Stern}, {Petty}, {Assef}, {Bussmann}, {Comerford},
  {Cutri}, {Evans}, {Griffith}, {Jarrett}, {Lake}, {Lonsdale}, {Rho},
  {Stanford}, {Weiner}, {Wright}, \& {Yan}}]{Wu2012}
{Wu}, J., {Tsai}, C.-W., {Sayers}, J., {et~al.} 2012, \apj, 756, 96

\bibitem[{{Y{\`e}che} {et~al.}(2010){Y{\`e}che}, {Petitjean}, {Rich},
  {Aubourg}, {Busca}, {Hamilton}, {Le Goff}, {Paris}, {Peirani}, {Pichon},
  {Rollinde}, \& {Vargas-Maga{\~n}a}}]{Yeche2010}
{Y{\`e}che}, C., {Petitjean}, P., {Rich}, J., {et~al.} 2010, \aap, 523, A14

\bibitem[{{Young}(2000)}]{Young2000}
{Young}, S. 2000, \mnras, 312, 567

\bibitem[{{Young} {et~al.}(2007){Young}, {Axon}, {Robinson}, {Hough}, \&
  {Smith}}]{Young2007}
{Young}, S., {Axon}, D.~J., {Robinson}, A., {Hough}, J.~H., \& {Smith}, J.~E.
  2007, \nat, 450, 74

\bibitem[{{Yuan} {et~al.}(2016){Yuan}, {Strauss}, \& {Zakamska}}]{Yuan2016}
{Yuan}, S., {Strauss}, M.~A., \& {Zakamska}, N.~L. 2016, \mnras, 462, 1603

\bibitem[{{Zakamska} \& {Greene}(2014)}]{Zakamska2014}
{Zakamska}, N.~L., \& {Greene}, J.~E. 2014, \mnras, 442, 784

\bibitem[{{Zakamska} {et~al.}(2003){Zakamska}, {Strauss}, {Krolik}, {Collinge},
  {Hall}, {Hao}, {Heckman}, {Ivezi{\'c}}, {Richards}, {Schlegel}, {Schneider},
  {Strateva}, {Vanden Berk}, {Anderson}, \& {Brinkmann}}]{Zakamska2003}
{Zakamska}, N.~L., {Strauss}, M.~A., {Krolik}, J.~H., {et~al.} 2003, \aj, 126,
  2125

\bibitem[{{Zakamska} {et~al.}(2005){Zakamska}, {Schmidt}, {Smith}, {Strauss},
  {Krolik}, {Hall}, {Richards}, {Schneider}, {Brinkmann}, \&
  {Szokoly}}]{Zakamska2005}
{Zakamska}, N.~L., {Schmidt}, G.~D., {Smith}, P.~S., {et~al.} 2005, \aj, 129,
  1212

\bibitem[{{Zakamska} {et~al.}(2006){Zakamska}, {Strauss}, {Krolik}, {Ridgway},
  {Schmidt}, {Smith}, {Heckman}, {Schneider}, {Hao}, \&
  {Brinkmann}}]{Zakamska2006}
{Zakamska}, N.~L., {Strauss}, M.~A., {Krolik}, J.~H., {et~al.} 2006, \aj, 132,
  1496

\bibitem[{{Zakamska} {et~al.}(2016){Zakamska}, {Hamann}, {P{\^a}ris}, {Brandt},
  {Greene}, {Strauss}, {Villforth}, {Wylezalek}, {Alexandroff}, \&
  {Ross}}]{Zakamska2016}
{Zakamska}, N.~L., {Hamann}, F., {P{\^a}ris}, I., {et~al.} 2016, \mnras, 459,
  3144

\bibitem[{{Zubko} \& {Laor}(2000)}]{Zubko2000}
{Zubko}, V.~G., \& {Laor}, A. 2000, The Astrophysical Journal Supplement
  Series, 128, 245

\end{thebibliography}
\bibliographystyle{aasjournal}

\appendix
\section{Additional figures}
\label{sec:app}

These figures show spectra and polarization data corresponding to
Figures \ref{fig:pol} and 
\ref{fig:velocity} for the four objects with lower S/N.  

\begin{figure*}
\includegraphics[scale=0.88,trim= 12 10 20 30,clip=true]{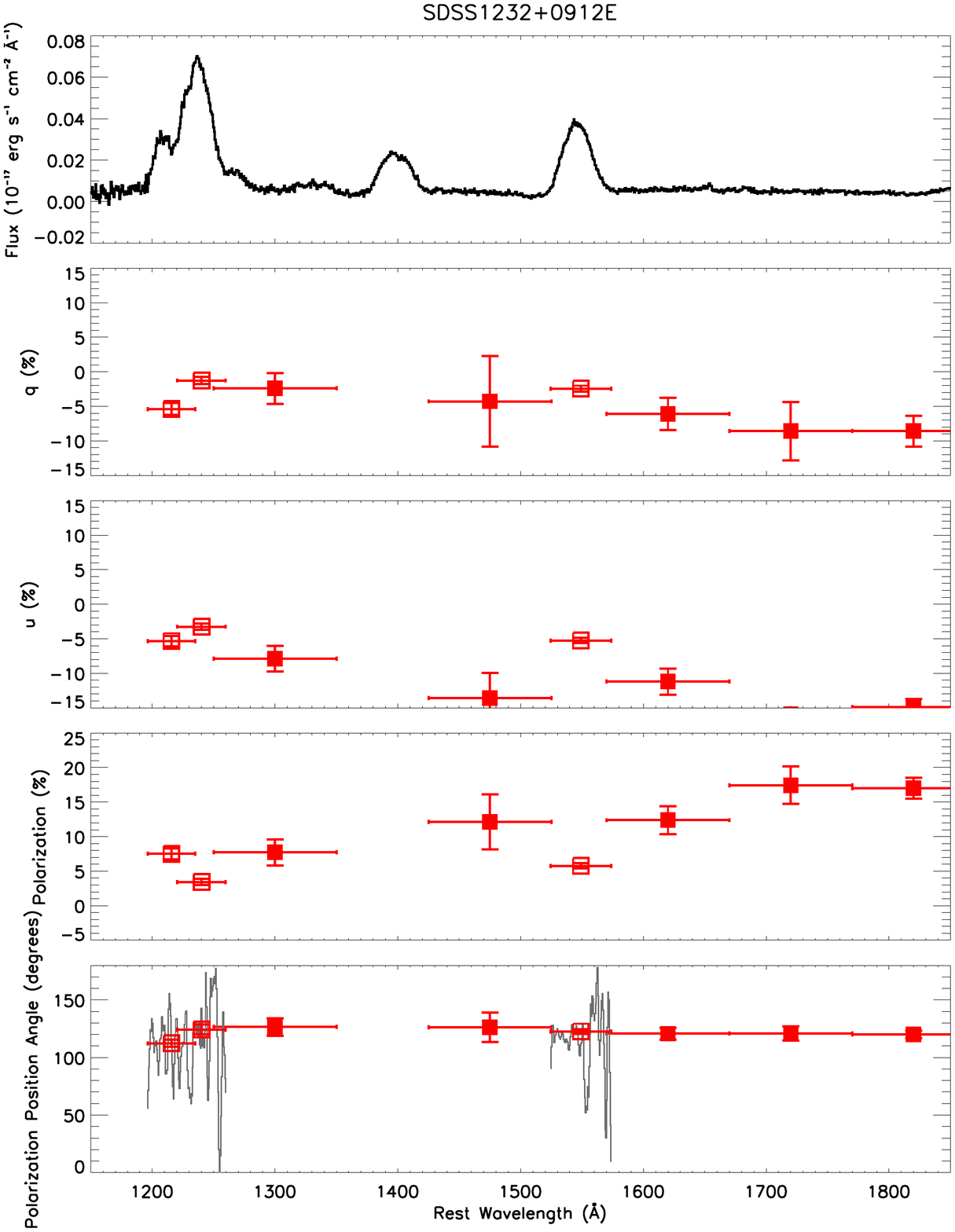}
\caption{\small LRISp spectra of our remaining targets (see Figure \ref{fig:pol}).}
\label{fig:pol_app1}
\end{figure*}

\begin{figure*}
\includegraphics[scale=0.88,trim= 12 10 20 30,clip=true]{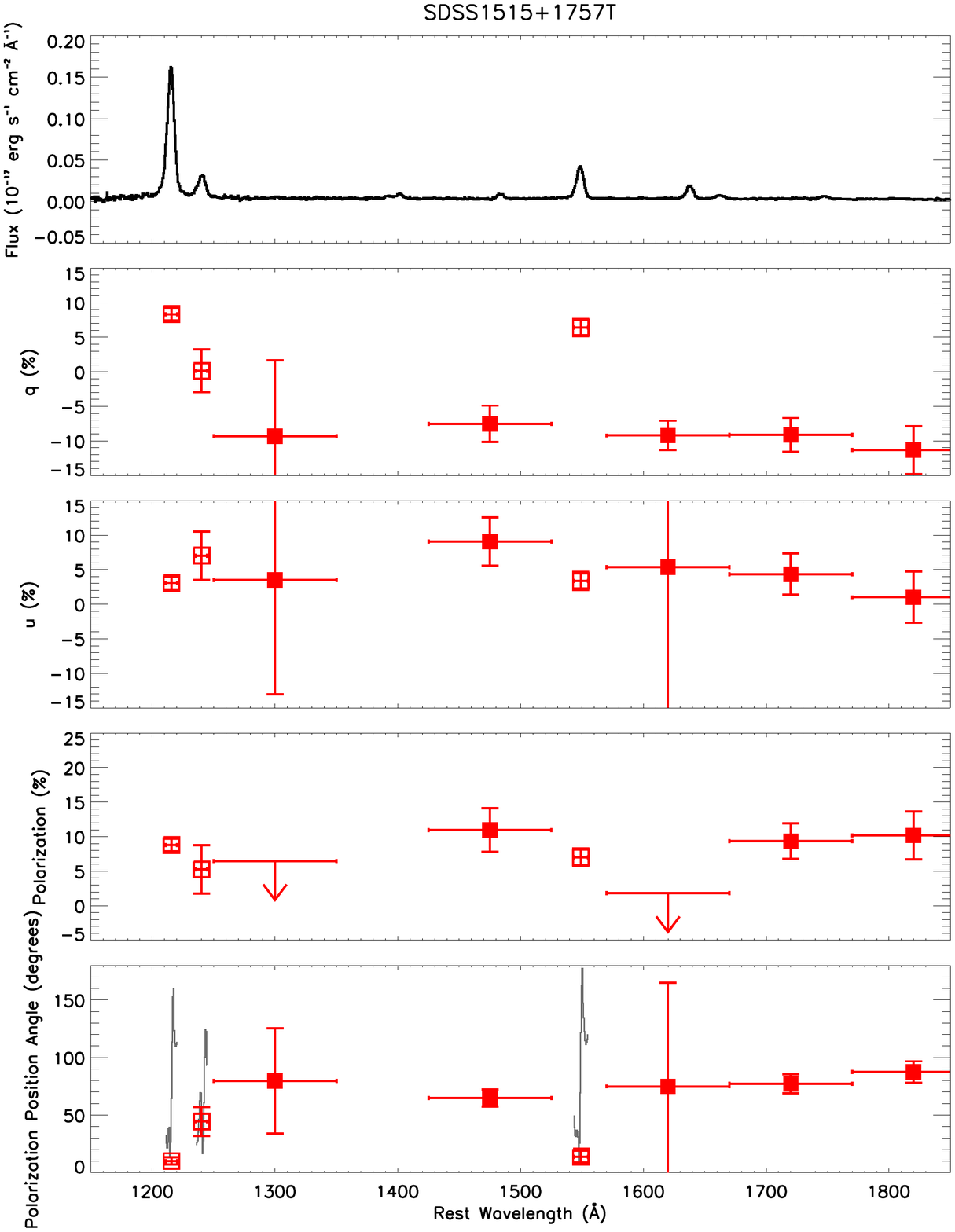}
\caption{\small continued}
\label{fig:pol_app2}
\end{figure*}

\begin{figure*}
\includegraphics[scale=0.88,trim= 12 10 20 30,clip=true]{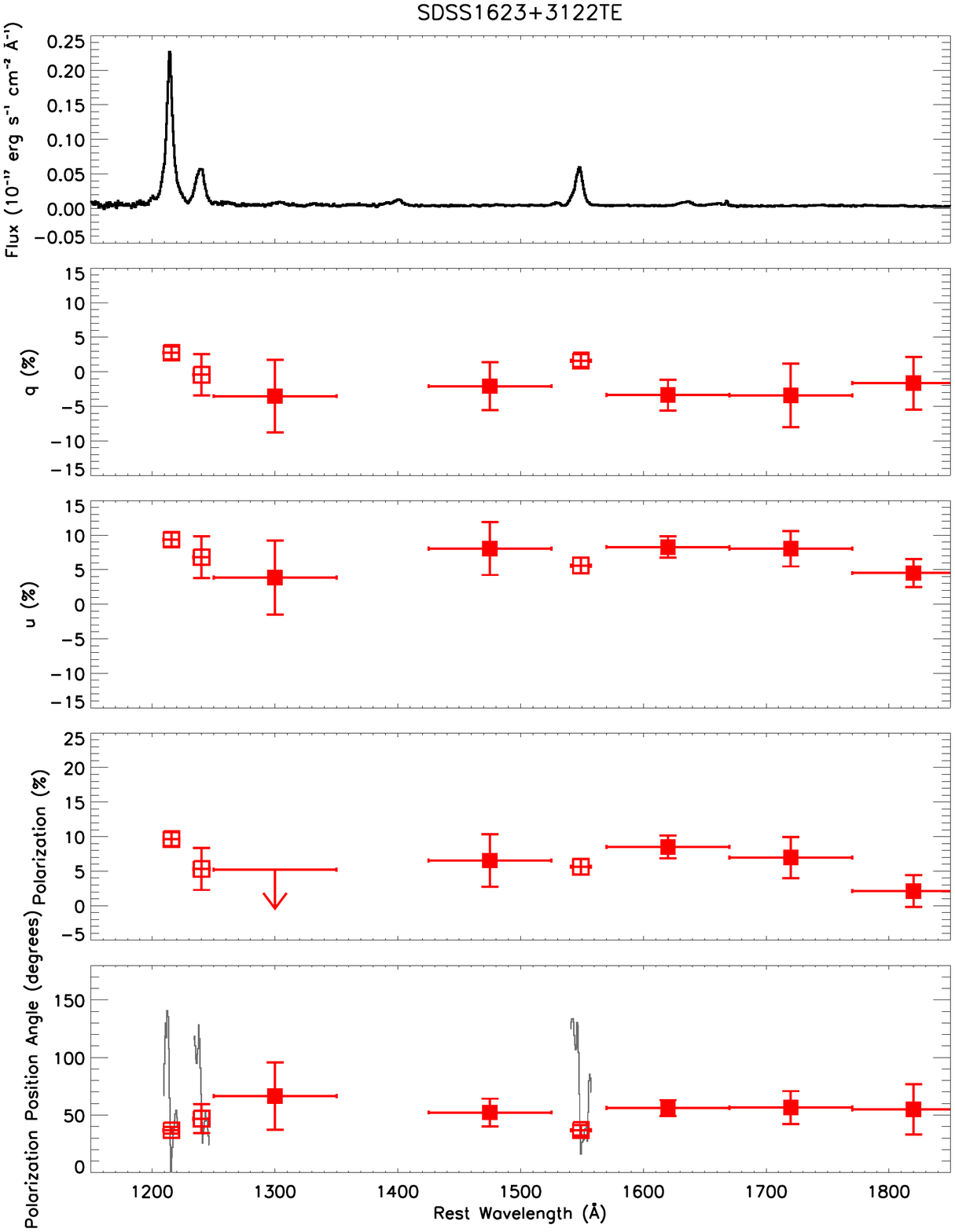}
\caption{\small continued}
\label{fig:pol_app3}
\end{figure*}

\begin{figure*}
\includegraphics[scale=0.88,trim= 12 10 20 30,clip=true]{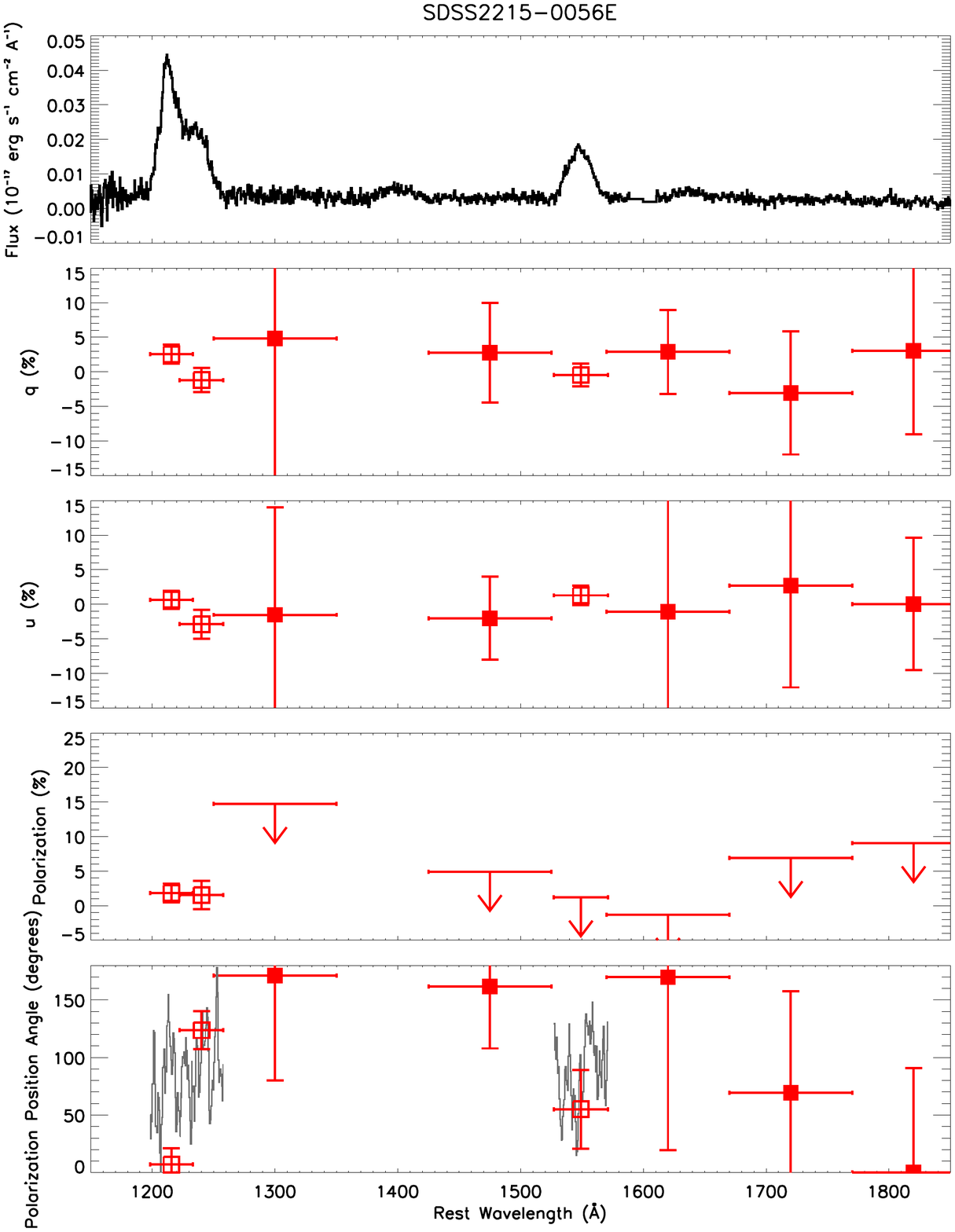}
\caption{\small continued}
\label{fig:pol_app4}
\end{figure*}

\begin{figure*}
\includegraphics[scale=0.7,trim=32 70 10 120,clip=true]{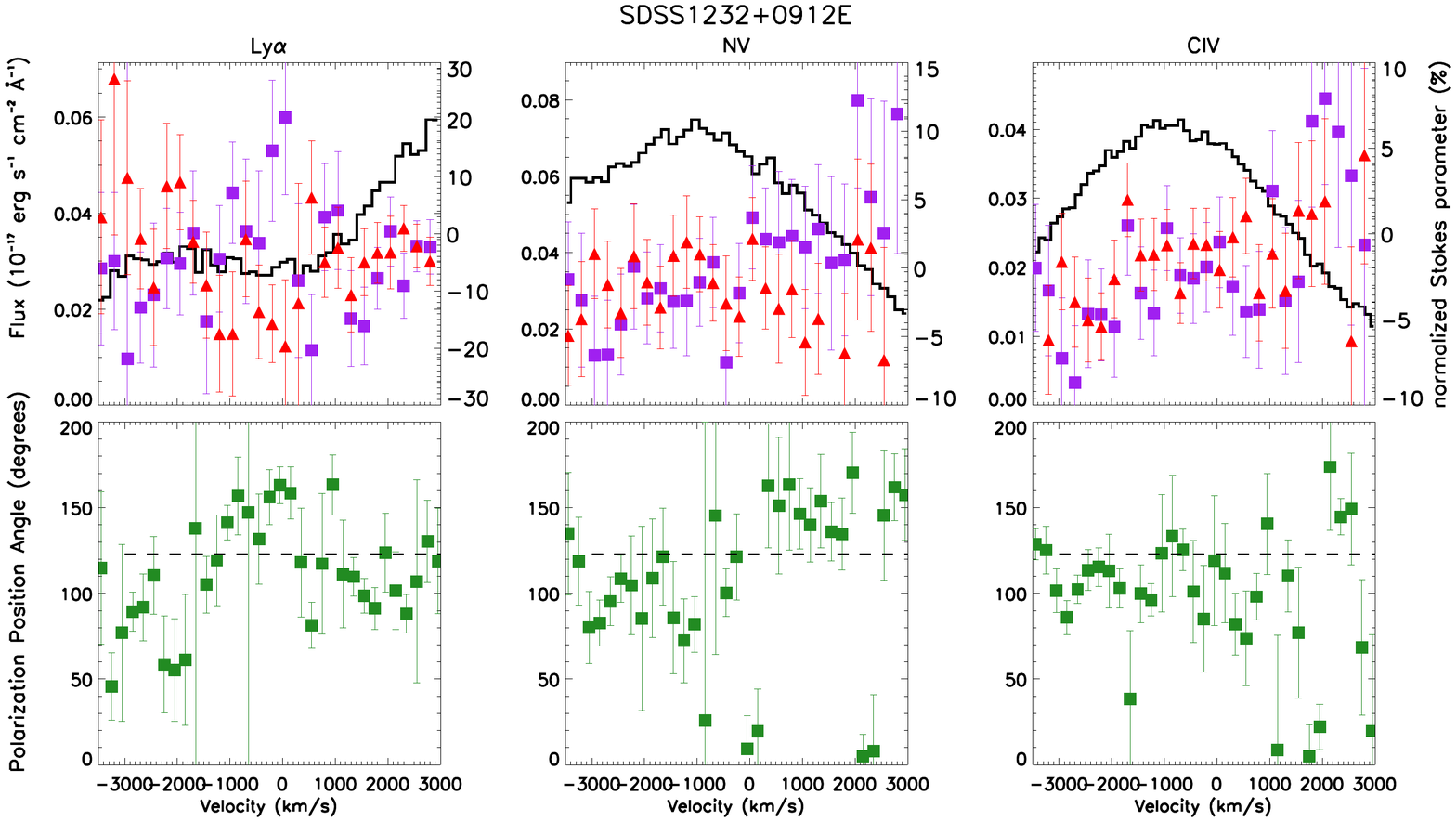}
\includegraphics[scale=0.7,trim=32 70 10 120,clip=true]{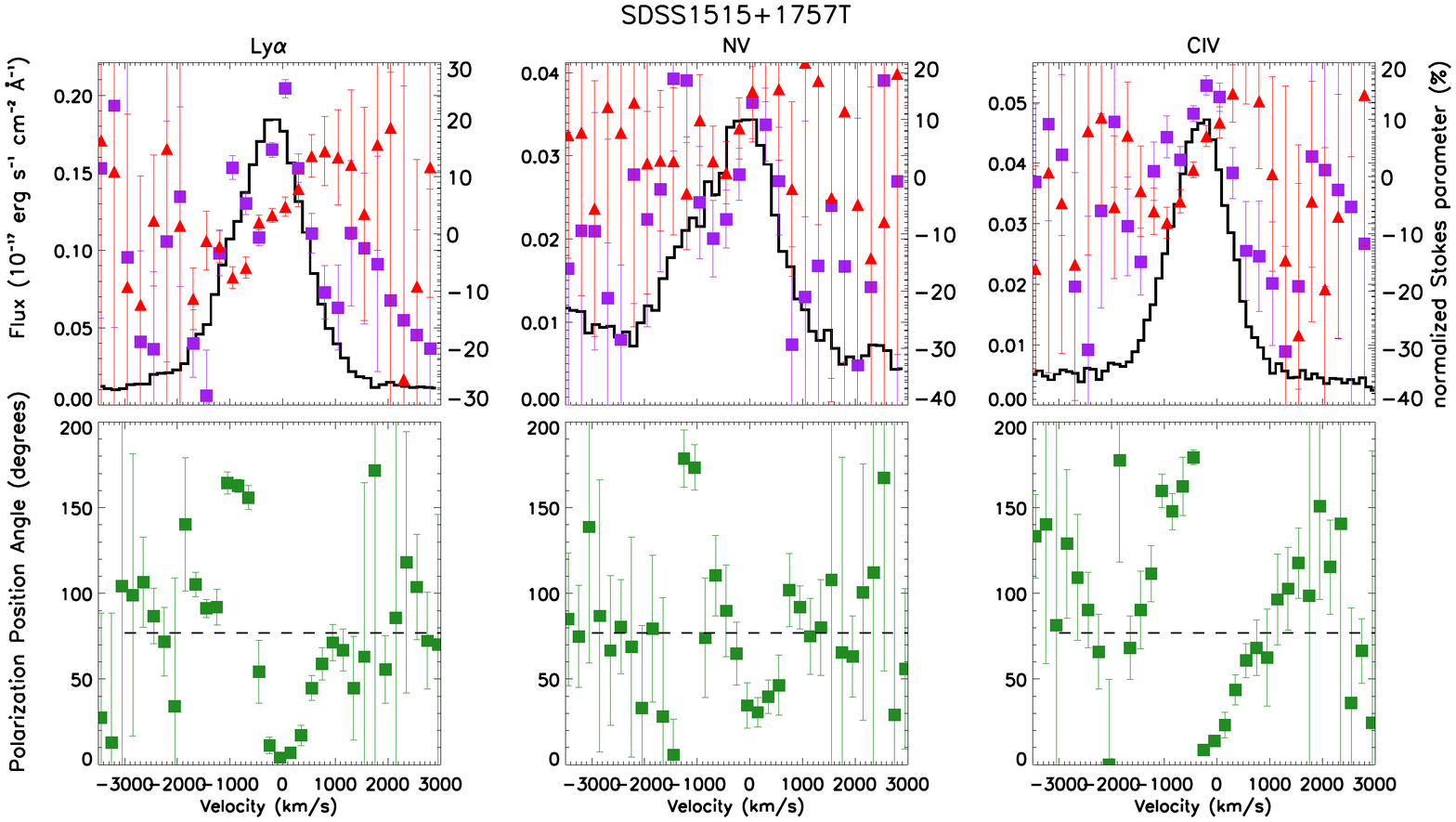}
\caption{\small Emission-line spectra of remaining targets (see Figure \ref{fig:velocity}).}
\label{fig:velocity_app1}
\end{figure*}

\begin{figure*}
\includegraphics[scale=0.7,trim=32 70 10 120,clip=true]{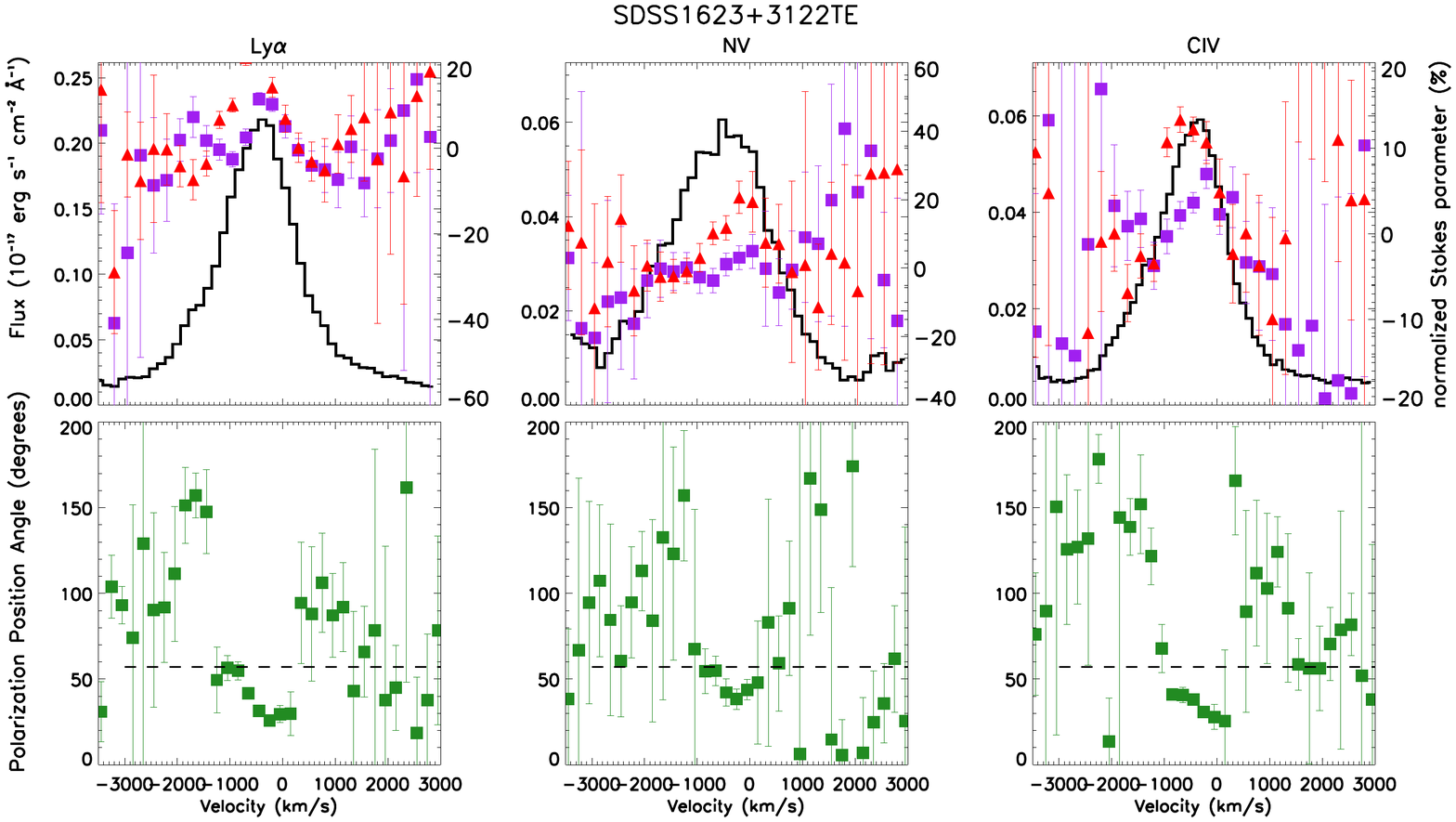}
\includegraphics[scale=0.7,trim=32 70 10 120,clip=true]{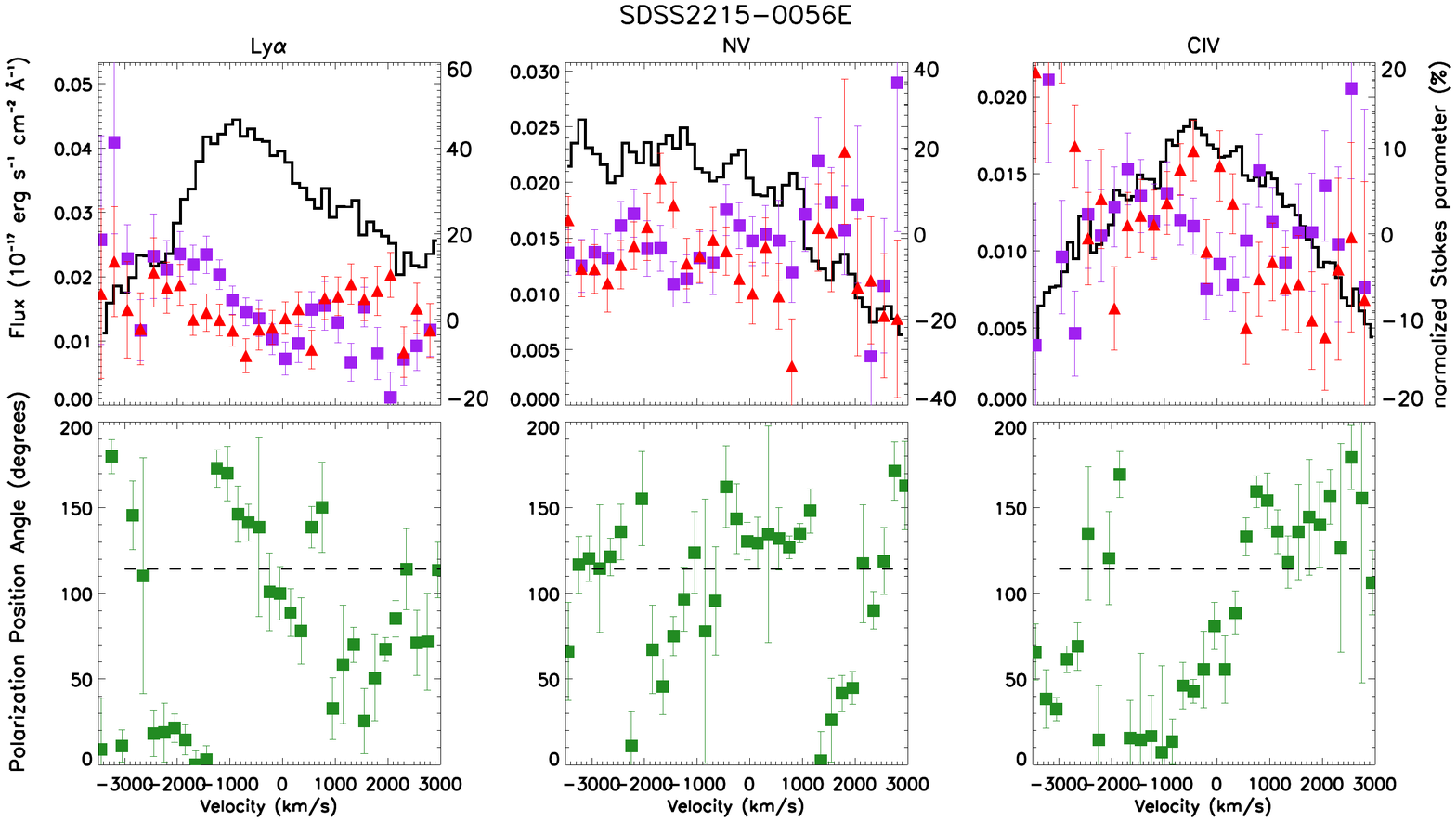}
\caption{\small continued}
\label{fig:velocity_app2}
\end{figure*}

\label{lastpage}

\end{document}